\documentclass[12pt]{article}

\usepackage{amsfonts,amsmath}
\usepackage{amssymb}
\usepackage{ytableau}
\usepackage{booktabs}
\usepackage{mathtools} % smashoperator for integrals
\usepackage{tikz-cd} % это тоже для диаграмм

\usepackage{pgfplots}
\pgfplotsset{compat=1.5}
\usepgfplotslibrary{fillbetween} 
\usepackage{tikz}
\usetikzlibrary{arrows.meta,bending,patterns,angles,calc}
\usetikzlibrary{patterns.meta}
\usepackage{siunitx}
\usetikzlibrary{decorations.shapes}
\tikzset{decorate sep/.style 2 args=
{decorate,decoration={shape backgrounds,shape=circle,shape size=#1,shape sep=#2}}}

\pgfplotsset{ignore zero/.style={%
  #1ticklabel={\ifdim\tick pt=0pt \else\pgfmathprintnumber{\tick}\fi}
}} %this is to have only one 0 on plot 

\newcommand{\dr}{{{\rm d}}}

\renewcommand{\theequation}{\thesection.\arabic{equation}}

\makeatletter \@addtoreset{equation}{section} \makeatother

{\vspace{3mm} }

\def\al{\alpha}

\def\*{\star}
\def\e{\mathbf{e}}
\def\E2{\mathbf{E}}
\def\w{\mathbf{w}}

\def\rmx{\mathrm{x}}
\def\rmz{\mathrm{z}}

\newcommand{\be}{\begin{equation}}
\newcommand{\ee}{\end{equation}}
\newcommand{\bee}{\begin{eqnarray}}
\newcommand{\beee}{\begin{array}}
\newcommand{\eee}{\end{eqnarray}}
\newcommand{\eeee}{\end{array}}
\newcommand{\gb}{\beta}
\newcommand{\gga}{\gamma}

\newcommand{\gd}{\delta}

\newcommand{\gep}{\epsilon}

\newcommand{\gs}{\sigma}
\newcommand{\go}{\omega}
\newcommand{\gO}{\Omega}

\newcommand{\dal}{\dot \alpha}
\newcommand{\dgb}{\dot \beta}
\newcommand{\dgga}{\dot \gamma}

\newcommand{\nn}{\nonumber}

\newcommand{\p}{\partial}

\newcommand{\ff}{\frac}
\newcommand{\rom}[1]{\uppercase

\expandafter{\romannumeral #1\relax}}

\begin{document}
    
\begin{flushright}
FIAN/TD/18-2025\\
\end{flushright}

\vspace{0.5cm}
\begin{center}
{\large\bf On symmetry breaking in the self-dual higher-spin theory}

\vspace{1 cm}

\textbf{V.E.~Didenko$^{1}$ and I.S.~Faliakhov$^{2}$}\\

\vspace{1 cm}

\textbf{}\textbf{}\\
 \vspace{0.5cm}
 \textit{$^{1}$I.E. Tamm Department of Theoretical Physics,
Lebedev Physical Institute,}\\
 \textit{ Leninsky prospect 53, 119991, Moscow, Russia }\\
 \vspace{0.5cm}
 {\it
			$^2$Moscow Institute of Physics and Technology,\\
			Institutsky lane 9, 141700, Dolgoprudny, Moscow region, Russia}

\par\end{center}

\begin{center}
\vspace{0.6cm}
e-mails: didenko@lpi.ru, faliakhov.is@phystech.edu \\
\par\end{center}

\vspace{0.4cm}

\begin{abstract}
\noindent We explore the symmetry-broken phase of the self-dual (chiral) sector of higher-spin theory in four dimensions. To that end, we construct a two-parameter vacuum that breaks the AdS symmetry but remains symmetric under the leftover Poincar\'{e} algebra in three dimensions. The vacuum non-zero fields include spin-two AdS frame fields and a scalar, which has a profile that extends along the AdS radial direction. The two free parameters correspond to two scalar branches of conformal dimensions $\Delta=1$ and $\Delta=2$. Focusing on the $\Delta=1$ branch, we analyze the dynamics of free fields around this vacuum and examine its holographic dual. 
We observe that certain higher spin states decouple in the broken phase. This is illustrated by  a set of gauge fluctuations, which acquire no source from higher-spin currents, leading to their complete decoupling, except for the gauge field associated with spin one. The dual higher-spin currents appear to be disentangled from the gauge fields and generally do not conserve; however, their lower-spin components with helicities $s = -1, 0$, and $\pm 1/2$ remain unaffected by the symmetry breaking. Notably, the helicity $s=+1$ current, while deformed,  remains conserved.

\end{abstract}

\newpage
\tableofcontents
\newpage

\section{Introduction}
Unitary higher-spin gauge theories (HS) in flat spacetime face significant challenges due to various no-go theorems \cite{Weinberg:1964ew, Coleman:1967ad, Aragone:1979hx}; see also \cite{Taronna:2017wbx, Roiban:2017iqg} for a more recent account and \cite{Bekaert:2010hw} for review. However, introducing a nonzero cosmological constant $\Lambda$ bypasses these flat-space restrictions and allows for higher-spin interactions, as demonstrated long ago by Fradkin and Vasiliev at the cubic level \cite{Fradkin:1986qy, Fradkin:1987ks}. These interaction vertices involve negative powers of $\Lambda$, which renders the straightforward limit $\Lambda\to 0$ meaningless.\footnote{One can make sense of this limit by $\Lambda$-rescaling HS fields, but the no-go's still prevent local interactions at higher orders. Notably, the self-dual interactions terminate at the cubic order in the light-cone frame as follows from Metsaev's analysis \cite{Metsaev:1991mt, Metsaev:PhD}; see also \cite{Ponomarev:2017nrr}. } Cubic HS interactions have been extensively studied in the literature; see \cite{Bekaert:2022poo} for a substantial list of references and \cite{Buchbinder:2021ite, Buchbinder:2022kzl, Tatarenko:2024csa, Zinoviev:2024eto} for more recent results, but research on higher-order interactions in AdS remains limited \cite{Bekaert:2015tva, Neiman:2023orj, Gelfond:2023fwe}. Nevertheless, it has long been established that the cubic Fradkin-Vasiliev interactions can be completed to all orders in four dimensions at the level of classical equations of motion \cite{Vasiliev:1992av}. This fundamental result introduces several important concepts:

\begin{enumerate}
    \item {\bf The higher-spin algebra}, in which  fields take values. 
    
    In four dimensions, the HS algebra is particularly straightforward to realize via the Moyal algebra of commuting variables $Y_A=(y_{\alpha}, \bar y_{\dal})$, where $\alpha$ and $\dot\alpha$ are $sl(2, \mathrm{C})$ indices that take on two values. The product in this algebra is defined as follows:
    \begin{equation}\label{Moyal}
        f(y, \bar y)*g(y, \bar y)=\int\limits_{U, V}  f(y+u, \bar y+\bar u)g(y+v, \bar y+\bar v)e^{iu_{\al}v^{\al}+i\bar u_{\dal}\bar v^{\dal}}\,,
    \end{equation}
    where the integration goes over $U=(u, \bar u)$ and $V=(v, \bar v)$ with the measure chosen such that $1*1=1$.  
    \item {\bf Deformation procedure:} This procedure is based on the unfolded formulation \cite{Vasiliev:1988sa} (for applications and quantum extension, see \cite{Misuna:2022cma, Misuna:2024ccj, Misuna:2024dlx}  ). It suggests that a convenient set of fields consists of gauge connections represented as space-time 1-forms $\omega(Y|x)=\dr x^{\mu}\omega_{\mu}(Y|x)$ and 0-forms $C(Y|x)$. The latter can be viewed as HS curvatures for the connections $\omega$ at the linearized level. Both $\omega$ and $C$ take values in the HS algebra. The deformation represents an action of the HS algebra and can be schematically expressed as:
     \begin{subequations}\label{unfld}
     \begin{align}
        &\dr_x\go=\sum_{n=0}^{\infty}\sum_{i+j=n}\eta^i\bar\eta^j\mathcal{V}_{i,j}(\go, \go, C^n)\,,\\
        &\dr_x C=\sum_{n=1}^{\infty}\sum_{i+j=n-1}\eta^i\bar\eta^j\Upsilon_{i,j}(\go, C^n)\,,
    \end{align}
    \end{subequations}  
    where $\mathcal{V}$ and $\Upsilon$ are polylinear functionals of $\omega$ and $C$. For example, we define $\mathcal{V}(\go, \go, C^2):=\mathcal{V}(\go, \go, C, C)$, which is linear in each argument. A constant $\eta$ is an arbitrary phase parameter of the four-dimensional theory with $\bar\eta$ being its complex conjugate. The parameter breaks parity unless $\eta=1$ or $\eta=i$. We refer to the functionals $\mathcal{V}$ and $\Upsilon$ as 1-form and 0-form vertices, respectively. Their explicit forms arise from the consistency $\dr_x^2=0$ and from the {\it initial data} corresponding to the non-deformed HS algebra action at the lowest order. Specifically, 
    \begin{equation}
        \mathcal{V}_{0,0}(\omega, \omega)=-\omega*\omega\,,\qquad \Upsilon_{0,0}(\go, C)=-\omega*C+C*\pi(\omega)\,,
    \end{equation}
    where $\pi f(y,\bar y):=f(-y, \bar y)$ is a certain automorphism of the HS algebra that ensures the free equations following from \eqref{unfld} align with those of Fronsdal \cite{Fronsdal:1978rb}.

    \item {\bf Vasiliev's generating equations:} These equations \cite{Vasiliev:1992av} allow for a systematic reconstruction of $\mathcal{V}$ and $\Upsilon$ from \eqref{unfld} up to a freedom in field redefinition; for reviews, see \cite{Vasiliev:1999ba, Bekaert:2004qos, Didenko:2014dwa}.  
   \end{enumerate}
Despite Vasiliev's approach providing access to any order of HS interactions, several technical and conceptual challenges remain. On the technical side, reconstructing the $C^3$ vertices and higher becomes increasingly complicated, see  e.g., \cite{Gelfond:2021two} and \cite{Gelfond:2023fwe}. More importantly, there is a natural freedom in field representatives, which means that the appropriate field variables corresponding to minimal (in terms of derivatives) interactions are not fully understood. This issue is known as the locality problem \cite{Prokushkin:1998bq, Giombi:2009wh, Boulanger:2015ova, Vasiliev:2016xui} and is currently the focus of intense research; for quite a  recent work in this direction, see \cite{Vasiliev:2023yzx, Kirakosiants:2025gpd}.   

The locality of the theory in AdS beyond the cubic order remains uncertain. Arguments based on HS holography \cite{Klebanov:2002ja, Leigh:2003gk, Sezgin:2003pt} suggest that it is not local, particularly starting from the quartic order \cite{Bekaert:2015tva, Sleight:2017pcz}. These arguments indicate a wild nonlocality of the form $1/\Box$, which, if accurate, would pose challenges for HS theories in AdS, similar to those observed in the Minkowski case. 

On the other hand, direct calculations from the bulk \cite{Gelfond:2023fwe} indicate a moderate nonlocality at the quartic order, which might be in tension with holographic expectations. However, let us emphasize that the holographic locality argument heavily depends on the assumption that the HS algebra remains undeformed at the boundary in interactions. Whether this assumption holds true is an open question.

\subsection{Holomorphic sector}
The locality problem becomes significantly simpler in the closed sector of the theory, which can be singled out by setting $\eta=0\ (\bar\eta=0)$, corresponding to the (anti)self-dual HS theory \cite{Vasiliev:1992av}. This sector is called  (anti)holomorphic, also known as (anti)chiral. In the system \eqref{unfld}, the relevant vertices are $\mathcal{V}_{0,n}$ and $\Upsilon_{0,n}$, or alternatively, $\mathcal{V}_{n,0}$ and $\Upsilon_{n,0}$. We refer to these as (anti)holomorphic vertices. It is worth emphasizing that relaxing the complex-conjugation $\bar\eta=\eta^*$ is not compatible with the reality conditions for HS fields (these can be found in, e.g., \cite{Vasiliev:1999ba}). Therefore, the holomorphic sector is essentially complex.    

In the self-dual case $\eta=0$, equations \eqref{unfld} simplify to the form
\begin{subequations}\label{unfld:hol}
     \begin{align}
        &\dr_x\go=\sum_{n=0}^{\infty}\mathcal{V}_n(\go, \go, C^n)\,,\label{unfld:hol-w}\\
        &\dr_x C=\sum_{n=1}^{\infty}\Upsilon_n(\go, C^n)\,,\label{unfld:hol-C}
    \end{align}
    \end{subequations} 
where we set $\bar\eta=1$ for convenience and use the simplified notation $\mathcal{V}_n:=\mathcal{V}_{0,n}$, $\Upsilon_n:=\Upsilon_{0,n}$. At the free level on AdS, the holomorphic sector also differs from the linearized full theory in that only HS Weyl tensors of helicities $s\leq -1$  source the HS gauge fields.\footnote{This may not be true on other backgrounds, such as the one we will consider in this paper.} In contrast, the full theory allows both helicities  $s\geq 1$ and $s\leq -1$  to source these fields. For other recent advances in chiral HS theory, see \cite{Tran:2022tft, Tsulaia:2022csz, Adamo:2022lah, Monteiro:2022xwq, Ponomarev:2024jyg, Serrani:2025owx, Bengtsson:2025dtk} and  also a notable 6D extension considered in \cite{Basile:2024raj}.

It has been conjectured in \cite{Gelfond:2018vmi} that the holomorphic vertices are local to all orders. This conjecture was tested up to the $C^3$ order \cite{Didenko:2020bxd} as new necessary tools were developed in a series of papers \cite{Gelfond:2018vmi, Didenko:2018fgx, Gelfond:2019tac, Didenko:2019xzz, Didenko:2022eso}. The all-order locality was confirmed in \cite{Didenko:2022qga}, where a Vasiliev-like generating system that reproduces equations \eqref{unfld:hol} was proposed. The structure of this system makes the holomorphic locality manifest in natural variables. Furthermore, the vertices can be explicitly extracted to all orders \cite{Didenko:2024zpd} (see also \cite{Sharapov:2022nps}, where local vertices were identified from the divergent homological perturbations). They exhibit what is known as projectively-compact spin locality \cite{Vasiliev:2022med}, which leads to space-time locality with a minimal number of derivatives.

\subsection{Higher-spin symmetry breaking}

The longstanding idea, originating from earlier works \cite{Gross:1987kza, Gross:1987ar, Gross:1988ue}, proposes the existence of a tensionless limit in superstring theory, which is believed to reveal a huge symmetry associated with massless gauge fields. In this context, the unbroken phase of superstring theory is viewed as a foundational hypothetical HS theory, while the superstring emerges from its symmetry-broken phase. These concepts have been further explored in the literature \cite{Sundborg:2000wp, Sezgin:2002rt, Bonelli:2003kh, Sagnotti:2003qa, Sagnotti:2011jdy}, but progress has been limited due to problematic interpretation in the flat superstring background, and the fact that string theory in AdS space has yet to be developed. For an interesting proposal of a HS theory that allegedly has room for stringy states, see \cite{Vasiliev:2018zer} and its subsequent development \cite{Tarusov:2025qfo}.

Another intriguing suggestion from \cite{Metsaev:1999kb} is that string theory might emerge at the boundary of AdS space in the HS broken phase\footnote{In the lower-dimensional topological case, a related analysis has been conducted in \cite{Chen:2025xlo}. Earlier analysis of the tensionless strings on orbifolds is available in \cite{Gaberdiel:2018rqv, Eberhardt:2018ouy, Eberhardt:2019ywk} }. In our paper, we aim to follow this line of reasoning, albeit in a simplified context that allows for a more detailed analysis. Specifically, we focus on the holomorphic HS sector, which is important because it is closed and yet describes local interactions in spacetime at all orders of perturbation theory. This makes it a valuable toy model for studying symmetry breaking. 

Our approach begins with the search for a holomorphic vacuum that mildly breaks AdS symmetry while preserving the Poincar\'{e} symmetry of its 3D boundary. The existence of such a vacuum has already been explored for the off-shell $d$-dimensional bosonic equations in \cite{Didenko:2023vna}, with a solution provided in \cite{Didenko:2023txr}; see also \cite{Aros:2017ror}. We revisit this problem within the purely holomorphic sector and demonstrate that the required vacuum, which satisfies equation \eqref{unfld:hol}, does indeed exist and has a schematic form: 
\begin{align}
    &\go_0=\Omega_{AdS}\,,\label{vac:ads}\\
    &C_0\quad\to\quad \phi=\nu_1 \rmz+\nu_2\rmz^2\,.\label{vac:scl}
\end{align}
The vacuum contains the standard AdS connection \eqref{vac:ads}. However, the global AdS symmetry is broken due to the presence of a nonzero matter expectation \eqref{vac:scl}, which is zero in the standard HS vacuum. In our case, the vacuum value includes only a specific scalar mode $\phi$ that depends on the inverse AdS radial distance $\rmz$, associated with the Poncar\'{e} metric (see eq. \eqref{Poincare}). The two coefficients, $\nu_1$ and $\nu_2$, can be arbitrary.   

Next, we further restrict ourselves to the case where $\nu_2 = 0$, allowing $\nu_1$ to remain arbitrary. We then study the linearized fluctuations around the vacuum \eqref{vac:ads}, \eqref{vac:scl}
\begin{equation}
    \go=\go_0+\w+\dots\,,\qquad C=C_0+\mathcal{C}+\dots\,.
\end{equation}
In the unbroken case with $C_0 = 0$, these fluctuations correspond to free massless fields of helicities $s\leq -1$ propagating on the AdS background. The boundary dual of these fluctuations leads to a set of conserved 3D currents. These currents typically source the boundary HS connections and take a remarkable form of a tensor product of two singletons. This fundamental result of Flato and Fronsdal \cite{Flato:1978qz} lays the groundwork for HS holography in \cite{Vasiliev:2012vf}; see also \cite{Diaz:2024kpr, Diaz:2024iuz}. We revisit the dual analysis in the HS broken case with $C_0\neq 0$, where all types of helicities are present and arrive at the following results:
\begin{itemize}
    \item A set of HS currents with helicities $s\geq -1$ appears to disentangle from HS connections, forming a closed 0-form module. The currents generally do not conserve, with the exception of the helicity $s=\pm 1$ current. 

    \item Naturally, the current equations are deformed by the symmetry-breaking parameter $\nu=\nu_1$, but this is not the case for helicities $s=-1, 0$ and $s=\pm\frac{1}{2}$, where they remain the same as in the unbroken case.

    \item We observe a set of HS connections that completely decouple from the system, having no currents as sources. The only exception is a spin one connection, which is sourced by the corresponding Maxwell current.  
\end{itemize}
Our findings highlight an intriguing aspect of the HS symmetry breaking mechanism. We demonstrate, loosely speaking, that higher spin currents are more likely to be broken compared to their lower spin counterparts. The breaking degree increases with the spin: the higher the spin, the more contributions it receives from smaller spins.   
  
This paper is structured as follows. Section \ref{sec:eqs} introduces the generating equations for HS self-dual interactions and presents the corresponding interaction vertices. Section \ref{sec:vac} proposes a vacuum state with broken HS symmetry. Section \ref{sec:lin} discusses the dynamics of free fields around this vacuum, while in Section \ref{sec:dual} we examine its holographic dual. Finally, we conclude in Section \ref{sec:conc}. Additionally, the paper includes five technical appendices for further reference.

\section{Equations of motion}\label{sec:eqs}
The unfolded form of the HS equations \eqref{unfld} arises order by order as a solution to the Vasiliev system \cite{Vasiliev:1992av}. In the self-dual case, where $\bar\eta=0$, it is convenient to use the generating equations proposed in \cite{Didenko:2022qga}\footnote{We expect that equations from \cite{Didenko:2022qga} should follow in a star-product reordering limit from those of Vasiliev but, strictly speaking, the equivalence of the two systems was established only for a few orders (see \cite{Didenko:2024zpd}). While it may seem unlikely, the possibility that they could diverge at sufficiently high orders was not ruled out.}. In particular, locality of HS interactions in the self-dual case is established in \cite{Didenko:2022qga} to all orders, and the corresponding vertices are available. 

To formulate the equations, we introduce an auxiliary spinor variable\footnote{The auxiliary spinor $z$ should not be confused with the Poincar\'{e} radial coordinate $\rmz$.} $z_{\al}=(z_1, z_2)$. This variable expands the original $Y$-space and allows us to introduce the field $W(z; y, \bar y)$, which is designed to capture nonlinear corrections in $\omega$ and $C$ through its dependence on $z$
\begin{equation}\label{W: decomp}
    W:=\go(y, \bar y)+W_1[\omega, C](z; y, \bar y)+W_2[\omega, C,C](z; y, \bar y)+\dots
\end{equation}
Notice that while $C$ can enter $W$ nonlinearly, the field $\omega$ being a 1-form enters $W$ strictly linearly. The equations that generate self-dual HS interactions take the following form  \cite{Didenko:2022qga}:  
\begin{subequations}\label{Eqs}
\begin{align}   
&\dr_x W+W*W=0\,,\label{Eq: dW}\\
&\dr_z W+\{W, \Lambda\}_*+\dr_x\Lambda=0\,,\label{Eq: dL}\\
&\dr_x C+\left(W(z'; y, \bar y)*C-C*W(z'; -y, \bar
y)\right)\Big|_{z'=-y}=0\,.\label{Eq: dC}
\end{align}
\end{subequations}
They contain two types of differentials: the space-time $\dr_x=\dr x^{a}\frac{\partial}{\partial x^{a}}$, and the auxiliary one $\dr_z=\dr z^{\al}\frac{\partial}{\partial z^{\al}}$. There is the auxiliary connection $\Lambda$ associated with the $\dr_z$-direction 
\be\label{Lambda}
\Lambda[C]=\dr z^{\al}z_{\al}\int_{0}^{1}d\tau\tau\, C(-\tau z,\bar
y)e^{i\tau zy}\,,
\ee
which satisfies 
\begin{equation}
    \label{LambdaProp}
    \dr_z \Lambda = C \ast \gamma, \quad \gamma = \frac{1}{2}\dr z^\beta \dr z_\beta\,  \exp{izy}\,.
\end{equation}
For brevity throughout the text, we frequently use the condensed notation for spinor contractions
\begin{equation}
    \xi\eta:=\xi_{\al}\eta^{\al}=-\eta\xi\,.
\end{equation}
The large star product is
\be\label{limst}
(f*g)(z, y; \bar y)= \int f\left(z+u', y+u; \bar y \right)\star
g\left(z-v,y+v+v'; \bar y \right)
\exp({iu_{\al}v^{\al}+iu'_{\al}v'^{\al}})\,,
\ee
where the measure in the integration over $u, u', v, v'$ is such that $1*1=1$, while $\star$ is the Moyal product with respect to variables $\bar y$
\begin{equation}\label{star:bar}
    f(\bar y)\star g(\bar y)=\int f(\bar y +\bar u)g(\bar y+\bar v)e^{i\bar u\bar v}\,.
\end{equation}
The following action can be easily derived from \eqref{limst} and \eqref{star:bar}:
\begin{subequations}\label{star:gen}
\begin{align}
&y_{\al}* =y_{\al}+i\ff{\p}{\p y^{\al}}-i\ff{\p}{\p z^{\al}}\,,\qquad z_{\al}* =z_{\al}+i\ff{\p}{\p
y^{\al}}\,,\\
&* y_{\al}=y_{\al}-i\ff{\p}{\p y^{\al}}-i\ff{\p}{\p z^{\al}}\,,\qquad * z_{\al}=z_{\al}+i\ff{\p}{\p
y^{\al}}\,,\\
&\bar y_{\dal}* =\bar y_{\dal}+i\ff{\p}{\p \bar y^{\dal}}\,,\qquad *
\bar y_{\dal}=\bar y_{\dal}-i\ff{\p}{\p \bar y^{\dal}}\,.
\end{align}
\end{subequations}
In particular, 
\begin{equation}
    [y_{\al}, y_{\gb}]_*=2i\gep_{\al\gb}\,,\quad [\bar y_{\dal}, \bar y_{\dgb}]_*=2i\gep_{\dal\dgb}\,,\quad [y_{\al}, z_{\gb}]_*=[\bar y_{\dal}, z_{\gb}]_*=[z_{\al}, z_{\gb}]_*=0\,.
\end{equation}
For $z$-independent functions the star product \eqref{limst} is just the Moyal product \eqref{Moyal}. 

Let us clarify the role of each equation from \eqref{Eqs} in relation to the unfolded equations \eqref{unfld:hol} they generate. 
\begin{itemize}
    \item Equation \eqref{Eq: dW}: When we substitute \eqref{W: decomp} into this equation, we arrive at \eqref{unfld:hol-w}.

    \item Equation \eqref{Eq: dL}: This equation specifies the necessary corrections in \eqref{W: decomp} and ensures that the  $z$-dependence is eliminated from \eqref{Eq: dW}. It can be solved order by order in $C$ in perturbation. This procedure features ambiguity in homogeneous solution at each order, manifesting field redefintion freedom 
    \begin{equation}
        W_{n}[\omega, C\dots C](z; Y)\to W_{n}[\omega, C\dots C](z; Y)+F_{n}[\omega, C\dots C](Y)\,,
    \end{equation}
    where $n$ denotes the order of perturbations in $C$. A convenient way to fix this freedom is by picking the physical field $\go(Y)$ at the zeroth order. This implies that
    \begin{equation}\label{W:can}
        W_{n}(0; y, \bar y)=0\,,\qquad n\geq 1\,.
    \end{equation}
The field frame specified in \eqref{W:can} is called {\it canonical}. 
       \item Equation \eqref{Eq: dC}: Upon substituting  $W$  (which solves \eqref{Eq: dL}), this equation reproduces \eqref{unfld:hol-C}.
       \item Dynamical content is collected in the 1-form $\go(y, \bar y)$ and the 0-form $C(y, \bar y)$ as follows: self-dual spin $S$ primaries are singled out by 
       \begin{subequations}\label{prime}
          \begin{align}
           &\left(y^{\al}\frac{\p}{\p y^{\al}}+\bar y^{\dal}\frac{\p}{\p\bar y^{\dal}}\right)\go=2(S-1)\go\,,\\
            &\left(y^{\al}\frac{\p}{\p y^{\al}}-\bar y^{\dal}\frac{\p}{\p\bar y^{\dal}}\right)C=2s\, C\,,\label{C:phys}
         \end{align}
      \end{subequations}
where $s=\pm S$ denotes helicity. Those components of $\go$ and $C$ that do not satisfy \eqref{prime} are expressed via derivatives of primaries on-shell.
\end{itemize}

It is important to emphasize that  $z'$  in \eqref{Eq: dC} is not a variable associated with the star product; rather, it is a parameter that replaces $z$. After completing the star product integration with respect to $Y$ in \eqref{Eq: dC}, we set  $z'=-y$. Finally, we should stress that equation \eqref{Eq: dC} is not independent; it arises as an integrability condition, specifically the consequence of  $\dr_z^2 = 0$  from \eqref{Eq: dL}. Consistency of \eqref{Eqs} is not trivial. It is based on the so-called projective identity \cite{Didenko:2022qga}. A direct check of consistency, if somewhat involved, is also available; see \cite{Korybut:2025vdn}.

\paragraph{Vertices} The system \eqref{Eqs} generates vertices on the right-hand sides of \eqref{unfld:hol}. The explicit form of these vertices, which corresponds to the minimal space-time local interaction, was previously determined in \cite{Didenko:2024zpd}. Below, we present the final result. The 1-form sector vertices read
\begin{equation}
    \label{verw}
    \begin{split}
        \mathcal{V}(&\omega, \omega, C^n) = \smashoperator{ \sum_{0\leqslant k_1 \leqslant k_2 \leqslant n} } \, (-)^{k_2 - k_1 + 1} \smashoperator{\int_{\mathcal{D}^{[k_1, k_2]}_{n+1} } } d\xi d\eta\, \times   \\
        & \times \int \frac{d^2 u d^2 u' d^2 v d^2 v'}{(2\pi)^4} \, e^{iuv + iu'v'}\Bigg( \prod_{j=1}^{k_1} \star C\Big(\xi_j v + \eta_j v', \bar y \Big) \Bigg) \star  \\
        &  \star (\partial_{\al})^n \omega\Big(\xi_{n+1} y + u - \frac{S^{[k_1, k_2]}_{n} }{2} \, v', \bar y \Big) \star \Bigg( \prod_{j=k_1+1}^{k_2} \star C\Big(\xi_j v + \eta_j v', \bar y \Big) \Bigg) \star\\
        & \star (\partial^{\al})^n \omega\Big(\eta_{n+1} y + u' + \frac{S^{[k_1, k_2]}_{n} }{2} \, v, \bar y \Big) \star \Bigg( \prod_{l=k_2+1}^{n} \star C\Big(\xi_l v + \eta_l v', \bar y \Big) \Bigg)\,,
    \end{split}
\end{equation}
where $\p_{\al}\go(a, \bar a):=\frac{\partial}{\partial a^{\al}}\go(a, \bar a)$, 
\begin{equation}
    S^{[k_1, k_2]}_n = 1 - \sum_{s=1}^{k_2} \xi_s + \sum_{s=k_2+1}^{n}\xi_s +
    \sum_{s=1}^{k_1}\eta_s -
    \sum_{s=k_1 + 1}^{n}\eta_s + \sum_{i<j}^{n}(\xi_i \eta_j - \xi_j
    \eta_i)\,,
\end{equation}
and the compact $(\xi, \eta)$-integration domain is 
\begin{equation}
    \label{eq:1-VertexDomain}
    \mathcal{D}^{[k_1, k_2]}_{n+1} =
    \begin{cases}
        \eta_1 + ... + \eta_{n+1} = 1, \quad \eta_i \geq 0 \, , \\
        \xi_1 + ... + \xi_{n+1} = 1, \quad \xi_i \geq 0 \, , \\
        \eta_i \xi_{i+1} - \eta_{i+1}\xi_i \geqslant 0, \quad i \in [0, k_1-1] \, ,  \\
        \eta_i \xi_{i+1} - \eta_{i+1}\xi_i \,  \boldsymbol{\leqslant }0, \quad i \in [k_1 + 1, k_2-1] \, , \\
        \eta_i \xi_{i+1} - \eta_{i+1}\xi_i \geqslant 0, \quad i \in [k_2 + 1, n]\,,
    \end{cases}
\end{equation}
where $\eta_0$ and $\xi_0$ should be identified with $\eta_{n+1}$ and $\xi_{n+1}$ correspondingly.

The vertices from the 0-form sector \eqref{unfld:hol-C} are
\begin{equation}
    \label{verC}
    \begin{split}
        &\Upsilon(\omega, C^n) =
    (-i)^{n-1}\sum_{k=0}^{n}(-)^{k+1}\smashoperator{\int_{\mathcal{D}_{n}^{[k]}} } d\xi d\eta \, 
    \int \frac{d^2 u d^2 v}{(2\pi)^2} \,e^{iuv}\times\\
    &\times\left({\prod_{j=1}^{k}}\star C(\xi_j y+\eta_j v,
    \bar y)\right)\star (y^{\al}\p_{\al})^{n-1}\go (S_{n}^{[k]}y+u,
    \bar y)\star \left({\prod_{j=k+1}^{n}}\star C(\xi_j y+\eta_j v,
\bar y)\right)
    \end{split}
\end{equation}
with 
\begin{equation}
    \label{eq:0-VertexDomain}
    \mathcal{D}^{[k]}_n =
    \begin{cases}
        \eta_1 + ... + \eta_{n} = 1\,,\quad\eta_i\geq 0\,, \\
        \xi_1 + ... + \xi_{n} = 1\,,\quad\xi_i\geq 0\,, \\
        \eta_i \xi_{i+1} - \eta_{i+1} \xi_{i} \leqslant 0 \,, \quad i \in [1 \,,\, k-1]\,, \\
        \eta_i \xi_{i+1} - \eta_{i+1} \xi_{i} \geqslant 0\,, \quad i \in [k+1\,,\,n-1]\,,
    \end{cases}
\end{equation}
and 
\begin{equation}\label{Snk}
    S^{[k]}_n = -\sum_{s=1}^{k} \xi_s + \sum_{s=k+1}^{n} \xi_s + \sum_{i<j}^{n}(\xi_i \eta_j - \xi_j \eta_i)\,.
\end{equation}
The integration domains \eqref{eq:0-VertexDomain} enjoy the following important property:
\be\label{domain}
\bigsqcup_{k-\textnormal{odd}}\mathcal{D}_n^{[k]}=\bigsqcup_{k-\textnormal{even}}\mathcal{D}_n^{[k]}\,,
\ee
which appears very helpful in practice. The symbol $\bigsqcup$ is understood as union up to a set of measure zero. Note that domains intersect only along the sets of measure zero, which can be omitted for integration domains. Since Eq. \eqref{domain} was not previously observed in \cite{Didenko:2024zpd}, we prove it in Appendix B.

\section{Vacuum solution}\label{sec:vac}
HS theories admit a natural vacuum, which is the AdS space-time. This is also the case with the self-dual sector. Specifically, equations \eqref{Eqs} are satisfied for
\begin{subequations}\label{vac:stnd}
    \begin{align}
        &C_0=0\,,\\
        &W_0:=\Omega(y, \bar y)=-\frac{i}{4}(\go_{\al\gb}y^\al y^\gb+\bar\go_{\dal\dgb}\bar y^{\dal}\bar y^{\dgb}+2\e_{\al\dgb}y^{\al}\bar y^{\dgb})\,,
    \end{align}
\end{subequations}
where 1-forms $\omega$ and $\e$ are associated respectively with the Lorentz connection and vierbein of the four-dimensional AdS. The corresponding Cartan equations following from \eqref{Eq: dW} read\footnote{We set the cosmological constant $\Lambda=-1$ for our convenience.}
\begin{subequations}\label{AdS:frame}
    \begin{align}
        &\dr_x\go_{\al\gb}+\go_{\al\gga}\wedge\go^{\gga}_{\hspace{5pt}\gb}+ \e_{\al\dgga}\wedge \e_{\gb}^{\hspace{5pt}\dgga}=0\,,\\
        &\dr_x\bar\go_{\dal\dgb}+\bar\go_{\dot{\alpha}\dot{\gamma}}\wedge \bar\go^{\dot{\gamma}}_{\hspace{5pt}\dot{\beta}}+\e_{\gamma\dot{\alpha}}\wedge \e^{\gamma}_{\hspace{5pt}\dot{\beta}}=0\,,\\
        &\dr_x\e_{\al\dal}+\go_{\al\gga}\wedge \e^{\gamma}_{\hspace{5pt}\dal}+\bar\go_{\dal\dgga}\wedge \e_{\alpha}^{\hspace{5pt}\dgga}=0\,.
    \end{align}
\end{subequations}
A convenient choice of the AdS coordinates in what follows is the Poincar\'{e} ones
\begin{equation}\label{Poincare}
    ds^2=\frac{1}{\rmz^2}(\dr\rmz^2+\dr\rmx_{\al\gb}\dr\rmx^{\al\gb})\,,
\end{equation}
where we introduced the boundary three-dimensional coordinates\footnote{To avoid confusion, we use $x$ to represent four-dimensional coordinates, while the three-dimensional slice is represented by $\rmx$.} 
\begin{equation}
    \rmx_{\al\gb}=\rmx_{\gb\al}\,,
\end{equation}
and the radial coordinate $\rmz$. The background connections \eqref{AdS:frame} can be chosen to be 
\begin{subequations}
    \begin{align}
        &\go_{\al\gb}=\frac{i}{2\rmz}\dr\rmx_{\al\gb}\,,\\
        &\bar\go_{\dal\dgb}=-\frac{i}{2\rmz}\dr\rmx_{\dal\dgb}\,,\\
        &\e_{\al\dgb}=\frac{i\epsilon_{\al\dgb}\,\dr\rmz - \dr\rmx_{\al\dgb}}{2\rmz}\,,\label{vierbein}
    \end{align}
\end{subequations}
where the 3+1 splitting is achieved by introducing 
\begin{equation}
    \gep_{\al\dgb}=-\gep_{\dgb\al}\,,\quad \gep^{\al\dgb}=-\gep^{\dgb\al}\,,\quad \gep_{\al\dgb}\gep^{\gb\dgb}=\delta_{\al}{}^{\gb}\,,\quad \gep_{\al\dal}\gep^{\al\dgb}=\delta_{\dal}{}^{\dgb}
\end{equation}
corresponding to a unit vector $v_{\al\dgb}=i\gep_{\al\dgb}$ along the $\rmz$ direction.\footnote{The factor of $i$ is due to the reality conditions $v^{\dagger}=v$.} We use the mixed epsilon to convert dotted and undotted indices according to the following convention:
\begin{align}
   &\bar y_{\al}:=\bar y^{\dgb}\gep_{\dgb\al}\,,\qquad y_{\dal}:=y^{\gb}\gep_{\gb\dal}\,.\\ 
   &\rmx_{\al\dgb}:=\rmx_{\al}{}^{\gb}\gep_{\gb\dgb}\,,\qquad \rmx_{\dal\dgb}=\rmx^{\al\gb}\gep_{\al\dal}\gep_{\gb\dgb}\,.
\end{align}
Note that both $\rmx_{\al\dgb}$ and $\rmx_{\dal\dgb}$ are not independent; rather, they are expressed via $\rmx_{\al\gb}$.  In these terms, the Poincar\'{e} vacuum connection amounts to 
\begin{equation}\label{Omega:vac}
    \Omega=\ff{1}{8\rmz}\dr\rmx^{\al\gb}(y+i\bar
y)^{2}_{\al\gb}+\ff{\dr\rmz}{4\rmz}y\bar y\,.
\end{equation}
Introducing 
\begin{equation}\label{y:pm}
    y^\pm_{\al}:=y_{\al}\pm i\bar y_{\al}\,,\qquad [y^\pm_{\al}, y^\pm_{\gb}]_*=0\,,
\end{equation}
we have various bilinears of $y^{\pm}$ to form the 3D conformal algebra $o(2,2)$. Specifically, 
\begin{subequations}\label{conf}
    \begin{align}
        &P_{\al\gb}=iy^+_{\al}y^{+}_{\gb}\,,\\
        &L_{\al\gb}=i(y^+_{\al}y^-_{\gb}+y^+_{\gb}y^-_{\al})\,,\\
        &K_{\al\gb}=iy^-_{\al}y^{-}_{\gb}\,,\\
        &D=\frac{i}{8}y^{-}_{\al}\bar y^{+\al}=-\frac{1}{4}y\bar y\,,\label{D}
    \end{align}
\end{subequations}
where $P$, $L$, $K$ and $D$ are the generators of translations, Lorentz boosts, special conformal transformations, and dilatation, respectively. The commutation relations read
\begin{subequations}   
    \begin{align}
        &[P_{\alpha\beta}, L_{\mu\nu}] = -4(\epsilon_{\alpha\nu}P_{\mu\beta} + \epsilon_{\beta\nu}P_{\mu\alpha} + \epsilon_{\alpha\mu}P_{\nu\beta} + \epsilon_{\beta\mu}P_{\nu\alpha})\,,\\
        &[L_{\alpha\beta},L_{\mu\nu}] = 4(\epsilon_{\alpha\mu}L_{\beta\nu} + \epsilon_{\beta\mu}L_{\alpha\nu} + \epsilon_{\alpha\nu}L_{\beta\mu} + \epsilon_{\beta\nu}L_{\alpha\mu})\,,\\
        &[K_{\alpha\beta}, P_{\mu\nu}] = -2(\epsilon_{\alpha\mu}L_{\beta\nu} + \epsilon_{\alpha\nu}L_{\beta\mu} + \epsilon_{\beta\mu}L_{\alpha\nu} + \epsilon_{\beta\nu}L_{\alpha\mu}) - 32(\epsilon_{\alpha\mu}\epsilon_{\beta\nu} + \epsilon_{\alpha\nu}\epsilon_{\beta\mu})D\,,\\
        &[D, P_{\alpha\beta}] = - P_{\alpha\beta}\,,\qquad [D, K_{\alpha\beta}] =  K_{\alpha\beta}\,,\\
        &[D, L] = [P,P] = [D,D] = [K,K] = 0\,.
    \end{align}
\end{subequations}

\subsection{Scalar vacuum}
Interestingly, the self-dual HS theory admits another simple vacuum, where the background space is still AdS given by \eqref{Omega:vac}, while all fields vanish except for the scalar that spreads along the direction $\rmz$
\begin{equation}\label{scalar}
    \phi=\nu_1\rmz+\nu_2\rmz^2\,,
\end{equation}
where $\nu_1$ and $\nu_2$ are arbitrary constants. Notice that the scalar satisfies the free equation of motion in AdS and represents an arbitrary mixture of its two conformal branches of conformal dimensions $\Delta=1$ and $\Delta=2$. As a solution of the generating equations \eqref{Eqs}, it has the following form: 
\begin{align}
    &W_0=\Omega+\nu_2\cdot \frac{i\rmz}{2}\dr \rmx^{\al\gb}z_{\al}z_{\gb}\int_0^1d\tau\tau (1-\tau)e^{i\tau zy^+}\,,\label{W:vac}\\
    &C_0=\nu_1\rmz\,\exp{y\bar y}+\nu_2\rmz^2\,(1+y\bar y)\exp{y\bar y}\,,\label{C:vac}
\end{align}
where $y^+$ is defined in \eqref{y:pm}  and $\Omega$ is given by \eqref{Omega:vac}. Notice that even though \eqref{W:vac} departs from \eqref{Omega:vac} when $\nu_2\neq 0$, it still describes the AdS background in the Poincar\'{e} frame since $W|_{z=0}=\Omega$. Another way of seeing this is by substituting \eqref{W:vac} into \eqref{Eq: dW}. The scalar \eqref{scalar} arises as a nonzero vacuum expectation of the Weyl module $C_0(y, \bar y)$ primary
\begin{equation}
    \phi(x):=C_0(0, 0|x)\,,
\end{equation}
while the rest of the components in $C$ are its on-shell derivatives. The dilatation operator \eqref{D} acts on \eqref{C:vac} in the twisted-adjoint representation, and for each branch, one finds
\begin{equation}
    \{D, C_{0}^{\Delta}\}_*=\Delta C_0^{\Delta}\,,
\end{equation}
where  
\begin{equation}\label{C:branches}
    C_{0}^{\Delta=1}=\nu_1\rmz\,\exp{y\bar y}\,,\qquad C_{0}^{\Delta=2}=\nu_2\rmz^2(1+y\bar y)\exp{y\bar y}\,.
\end{equation}
Notably, the two branches are related to each other under the following Poincar\'{e} algebra action
\begin{equation}\label{eq:2brnch}
    \nu_2 [L_{\alpha\beta}, \Lambda[C_0^{\Delta=1}]]_* = {\frac{2\nu_1}{\rmz}}[P_{\alpha\beta}, \Lambda[C_0^{\Delta=2}]]_*\,,
\end{equation}
where $\Lambda[C]$ was defined in \eqref{Lambda}. We leave the derivation of vacuum \eqref{W:vac}, \eqref{C:vac} to Appendix A. In what follows, we set 
\begin{equation}\label{nu}
    \nu_2=0\quad\text{and}\quad \nu_1:=\nu\,,
\end{equation}
postponing the general case for future investigation.   

\paragraph{Global symmetries} The new vacuum breaks the global HS symmetry to a subalgebra that includes the three-dimensional Poincar\'{e} space-time algebra, acting on the variables $\rmx_{\al\gb}$ for arbitrary values of $\nu_1$ and $\nu_2$ (see also \cite{Didenko:2023txr}). Although it may appear that equations \eqref{W:vac} and \eqref{C:vac} smoothly deform the standard HS vacuum described in \eqref{vac:stnd}, resulting in a deformation of the global HS symmetry, this is not the case. Instead, the symmetry is actually broken down. This happens because \eqref{C:vac} contains the following structure:
\begin{equation}\label{Fock}
    \mathcal{P}=4e^{y\bar y}\,,
\end{equation}
which is, on the one hand, an analytic function of $y$ and $\bar y$, but on the other hand acts as a projector in the star-product algebra
\begin{equation}
    \mathcal{P}*\mathcal{P}=\mathcal{P}
\end{equation}
that can be checked using \eqref{limst}. As a result, $\mathcal{P}$ projects away some states, acting as a Fock vacuum. Specifically,  
\begin{equation}\label{Fock:anih}
    y^+*\mathcal{P}=\mathcal{P}*y^-=0\,.
\end{equation}
Let us now determine the global symmetry of the vacuum \eqref{W:vac}, \eqref{C:vac} with $\nu_2=0$. The corresponding global symmetry parameter of eqs. \eqref{Eqs} satisfies (see \cite{Didenko:2023vna})
\begin{subequations}\label{glob}
\begin{align}
    &\dr_x\gep+[W_0, \gep]_*=0\,,\label{glob:W}\\
    &\dr_z\gep+[\Lambda[C_0], \gep]_*=0\,,\label{glob:L}
\end{align}
\end{subequations}
where $\Lambda[C_0]$ is defined in the vacuum \eqref{C:vac} via \eqref{Lambda}, and $W_0$ is given in \eqref{W:vac}. The global space-time symmetry $\gep$ can be found in perturbations in $\nu$ from \eqref{glob}
\begin{equation}\label{gep:pert}
    \gep=\gep_0+\nu\gep_1+\dots
\end{equation}
Naturally, one expects $\gep_0$ to be bilinear in $Y$'s and $z$-independent. It is not difficult to show that the following parameter solves \eqref{glob:W}:
\begin{equation}\label{eps0}
    \gep_0=\xi^{\al\gb}_{P}P_{\al\gb}+\xi^{\al\gb}_{L}L_{\al\gb}\,,
\end{equation}
where $P$ and $L$ are the Poincar\'{e} generators from \eqref{conf}, while $\xi^{\al\gb}_{L}$ are arbitrary constants associated with the $3d$ Lorentz boosts, and %{\color{red} It should be calculated below} 
\begin{equation}
    \xi^{\al\gb}_P=-\frac{3i}{4\rmz}\left(\xi_{0}^{\al\gb}+\xi_{L}^{\al\gga}\rmx_{\gga}{}^{\gb}+\xi_{L}^{\gb\gga}\rmx_{\gga}{}^{\al}\right)
\end{equation}
contains constant parameters $\xi_{0}^{\al\gb}$ associated with the $3d$ translations. The result presented in \eqref{eps0}, however, does not satisfy \eqref{glob:L}, because  $[\Lambda, \gep_0]_*\neq 0$. This induces a nonzero $O(\nu)$-correction to $\gep$ satisfying 
\begin{equation}
    \nu\dr_z\gep_1=-[\Lambda[C_0], \gep_0]_*\,.
\end{equation}
The latter could be solved using the standard contracting homotopy\footnote{Whenever consistent, equation of the form $\dr_{z}f(z)=\dr z^{\al}X_{\al}(z)$ can be solved by $f=z^{\al}\int_{0}^{1}d\tau X_{\al}(\tau z)$, known as contracting homotopy. \label{footnote}} with the result being 
\begin{equation}\label{gep1}
    \gep_1= 8\rmz\,\xi_{L}^{\al\gb}z_{\al}z_{\gb}\int_{0}^{1}d\tau\,\tau(1-\tau)e^{i\tau\,zy^+}\,.%\quad\text{\color{red}changed coefficient}
\end{equation}
It can be verified now by a straightforward computation that 
\begin{equation}
    [\gep_1, \Lambda[C_0]]_*=0\,.
\end{equation}
 Notice that the dependence on oscillators $z, y, \bar y$ in \eqref{gep1} is exactly the same as in the $\Delta=2$ connection corrections to $\Omega$ in \eqref{W:vac}. 
 This entails $\gep_1$ from \eqref{gep1} satisfies (see \eqref{eq:dw} and \eqref{w:corr} for a very similar analysis)
\begin{equation}
    \dr_x\gep_1+[W_0, \gep_1]_*=0\,.
\end{equation}
Therefore, the perturbation series \eqref{gep:pert} terminates at first order in $\nu$ that makes
\begin{equation}
    \gep=\gep_0 + 8\nu\rmz\,\xi_{L}^{\al\gb}z_{\al}z_{\gb}\int_{0}^{1}d\tau\,\tau(1-\tau)e^{i\tau\,zy^+}
\end{equation}
an exact solution of Eqs. \eqref{glob}. 
Thus, the leftover global symmetry is indeed generated by the boundary Poincar\'{e} algebra with parameters $\xi_L$ and $\xi_P$. It can be shown that no other generators from \eqref{conf} are the symmetries of the vacuum. Indeed, the full form of the global symmetry is determined perturbatively from knowing $\gep_0$ of \eqref{gep:pert}. The lowest order in $\nu$ consistency of Eq. \eqref{glob:L} implies
\begin{equation}\label{sym:cons}
    \gd_{\gep_0}C_0=-\gep_0*C_0+C_0*\pi(\gep_0)=0\,.
\end{equation}
It is not difficult to make sure that only the generators from \eqref{eps0} fulfill the condition \eqref{sym:cons}.

As a side remark, it is of interest to note  that corrections \eqref{w:corr} and \eqref{gep1} coincide. Technically, this occurs due to a nice relationship between the two vacuum branches \eqref{eq:2brnch} through the Poincar\'{e} algebra. This observation might have implications for the analysis of the $\Delta=2$ branch, which we leave for future investigation.

\section{Linearization}\label{sec:lin}
Having settled with the vacuum, one can address the problem of linear fluctuations about it. Our choice of interest is \eqref{nu} and, therefore, the vacuum fields are
\begin{subequations}\label{vac}
    \begin{align}
        &W_0:=\Omega\,,\\
        &C_0:=\nu\rmz\, e^{y\bar y}\,,\label{C0}\\
        &\Lambda_0:=\Lambda[C_0]=\nu\,\rmz\,\dr z^{\al}z_{\al}\int_{0}^{1}d\tau\tau e^{i\tau zy^+}\,.\label{Lambda0}
    \end{align}
\end{subequations}
Linearization of the unfolded HS equations \eqref{unfld:hol}
\begin{equation}\label{fields: lin}
    C=C_0+\mathcal{C}\,,\qquad \go=\Omega+\w
\end{equation}
amounts to 
\begin{subequations}\label{lin:unfld}
    \begin{align}
       &\dr_x\w+\{\Omega, \w\}_*=\sum_{n>0}\mathcal{V}_n(\Omega, \w, C_0^n)+\sum_{n>0}\mathcal{V}_n(\Omega, \Omega, \mathcal{C},C_0^{n-1})\,,\label{w:lin}\\
       &\dr_x\mathcal{C}+\w*C_0-C_0*\pi(\w)+\Omega*\mathcal{C}-\mathcal{C}*\pi(\Omega)
    =\sum_{n>1}\Upsilon_n(\w,C_0^{n})+\sum_{n>1}\Upsilon_{n+1}(\Omega, \mathcal{C},
    C_0^n)\,,\label{C:lin}
    \end{align}
\end{subequations}
where, abusing notation, we denote, say, $\mathcal{V}(\Omega, \w, C_0,\dots ,C_0):=\mathcal{V}_n(\Omega, \w, C_0^n)$. Equivalently, the right-hand sides of eqs. \eqref{lin:unfld} can be approached by discarding nonlinear terms in the generating system \eqref{Eqs} around \eqref{vac}. The decomposition takes the form 
\begin{subequations}\label{lin:gen}
    \begin{align}
        &\dr_x W+\{\Omega, W\}_*=0\,,\label{lin:dw}\\
        &\dr_z W+\{\Lambda_0, W\}_*+\{\Lambda, \Omega\}_*+\dr_x\Lambda=0\,,\label{lin: dL}\\
        &\dr_x \mathcal{C}+\Omega*\mathcal{C}-\mathcal{C}*\pi(\Omega)+\left(W(z'; y, \bar y)*C_0-C_0*W(z'; -y, \bar
y)\right)\Big|_{z'=-y}=0\,.
    \end{align}
\end{subequations}
Here, by $W$ and $\mathcal{C}$ we denote fluctuations around $\Omega$ and $C_0$, correspondingly, and 
\begin{equation}
    \pi(\Omega(y, \bar y))=\Omega(-y, \bar y)\,.
\end{equation}
As usually, the connection $\Lambda$ is expressed via the $z$-independent $\mathcal{C}(y, \bar y)$ as follows:
\begin{equation}
    \Lambda:=\Lambda[\mathcal{C}]= \dr z^{\alpha}z_{\alpha}\int_0^1 d\tau\ \tau\, \mathcal{C}(-\tau z, \bar y) e^{i\tau zy}\,.
\end{equation}
Although it may appear unnecessary due to the availability of all vertices in \eqref{lin:unfld} from \eqref{verw} and \eqref{verC}, analyzing the generating equations \eqref{lin:gen} reveals structures that are not immediately apparent from \eqref{verw} and \eqref{verC}. Therefore, we proceed using \eqref{lin:gen}. 

\subsection{Gauge fields}
Gauge fields describing linear fluctuations around the background $\Omega$ are collected in $\w=\w(y, \bar y)$ according to \eqref{fields: lin}. The same function of two variables $y$ and $\bar y $ should appear in the process of reconstruction of $z$ dependence using Eqs. \eqref{lin:gen}. An important comment in this regard is now in order. The standard case of the free HS fields propagating on AdS corresponds to $\nu=0$, where $C_0=\Lambda_0=0$. In this case, Eq. \eqref{lin: dL} lacks the second term. The equation restores the $z$-dependence of the field $W$ up to a freedom parameterized by the kernel of the operator $\dr_z$, which is simply $z$-independent functions. This way the physical field $\go(y, \bar y)=\Omega+\w$ appears at the zeroth order as a solution of $\dr_z W=0$, where the background part associated with $\Omega$ is conventionally separated. For $\nu\neq 0$ the role of the operator $\dr_z$ is replaced with 
\begin{equation}\label{D:oper}
    \mathcal{D}_{\nu}=\dr_z+\{\Lambda_0, \bullet\}_*\,,
\end{equation}
while the physical field, which we now denote with $\mathcal{W}$, satisfies 
\begin{equation}\label{coh}
    \mathcal{D}_{\nu}\mathcal{W}=0\,.
\end{equation}
Indeed, now that $\nu\neq 0$, the reconstruction of the field $W$ in terms of $\mathcal{C}$ (hidden in $\Lambda$) from \eqref{lin: dL} is carried out up to a solution of the homogeneous equation \eqref{coh}. Since the operator $\mathcal{D}_{\nu}$ contains manifestly variable $z$, solutions of \eqref{coh} are not generally $z$ independent. Nevertheless, the space of solutions of \eqref{coh} should be parameterized by an arbitrary function $\w(y, \bar y)$ in accordance with eqs. \eqref{unfld} which follow from the generating system \eqref{lin:gen}. 

Let us analyze Eq. \eqref{coh} under the assumption that $\mathcal{W}$ is analytic in $\nu$. In this case, one can elaborate on a perturbation theory in $\nu$.
\begin{equation}\label{nu:pert}
    \mathcal{W}=\mathcal{W}^{(0)}+\nu\mathcal{W}^{(1)}+\nu^2\mathcal{W}^{(2)}+\dots
\end{equation}
Feeding this into \eqref{coh}, at the lowest order, we have that 
\begin{equation}
    \dr_{z}\mathcal{W}^{(0)}=0\quad\Rightarrow\quad \mathcal{W}^{(0)}= \mathcal{W}^{(0)}(y, \bar y)\,.
\end{equation}
At the next $O(\nu)$ order, we further have 
\begin{equation}
    \nu\dr_z\mathcal{W}^{(1)}+\{\Lambda_0, \mathcal{W}^{(0)}\}_*=0\,.
\end{equation}
The latter equation is generally inconsistent unless 
\begin{equation}
    \dr_z\{\Lambda_0, \mathcal{W}^{(0)}\}_*=0\,,
\end{equation}
which, using \eqref{LambdaProp}, reduces to 
\begin{equation}\label{w0:cons}
 \mathcal{W}^{(0)}*C_0-C_0*\pi(\mathcal{W}^{(0)})=0\,. 
\end{equation}
The fact that $\mathcal{W}^{(0)}$ is not an arbitrary function of $y$ and $\bar y$ may seem to be in tension with $\w(y, \bar y)$ from \eqref{fields: lin} being arbitrary. Notice, however, that the latter should be encoded in the full solution of \eqref{coh} rather than its perturbative part. 

It turns out there is a $z$ independent solution of \eqref{coh} that contains part of the full degrees of freedom. It can be approached as follows. Using that $C_0$ is proportional to the Fock projector \eqref{Fock} and that 
\begin{equation}
    \pi(y^\pm)=-y^{\mp}
\end{equation}
it follows from \eqref{Fock:anih} that any function of the form 
\begin{equation}
    \mathcal{W}^{(0)}=\mathcal{W}^{(0)}(y^+)
\end{equation}
solves \eqref{w0:cons}. Additionally, such functions enjoy 
\begin{equation}
    \{\Lambda_0, \mathcal{W}^{(0)}(y^+)\}_*=0
\end{equation}
as follows after straightforward calculation. Therefore, for functions of $y^+$, the perturbation theory in $\nu$ terminates at the first step, as any function $\w=\w(y^+)$ satisfies 
\begin{equation}\label{w:phys}
    \mathcal{W}(y, \bar y):=\w(y^+)\,,\qquad \mathcal{D}_{\nu}\w(y^+)=0\,.
\end{equation}
We can conclude that $\w(y^+)$ represents physical fluctuations and we choose them as a particular solution for $\mathcal{W}$.  However, these functions do not encompass all degrees of freedom. Indeed, the physical degrees of freedom are packed into an arbitrary function $\w(y^+, y^-)$ instead of $\w(y^+)$. In the $\nu\neq 0$ case, the missing degrees of freedom are embedded in the solutions of \eqref{coh}, which typically depend on $z$. In our analysis of the broken phase, we focus on the $z$-independent degrees of freedom. These are functions of the form $\w(y^+)$. Below we briefly discuss how the rest of the degrees of freedom arise as well as exclude certain seemingly natural $z$-independent candidates.

\subsection{Unaccounted degrees of freedom}   It was already stated that Eq. \eqref{w:phys} is not a general analytic solution of the equation \eqref{coh}. For instance, functions of the form
\begin{equation}\label{exmpl}
    \mathcal{W}^{(0)}=y^+_{\al}*\phi^{\alpha\beta}(y, \bar y)*y_{\gb}^+\,,
\end{equation}
where $\phi$ is arbitrary, they still meet the integrability condition given by \eqref{w0:cons}. Since \eqref{exmpl} generally no longer anticommutes with $\Lambda_0$, the field $\mathcal{W}^{(1)}$ becomes dependent on $z$. This process continues beyond this point, leading to the generation of $\mathcal{W}^{(2)}$ and so forth.

One might wonder what the other solutions of \eqref{coh} are? In particular, there could be non-analytic solutions in $\nu$ for which the decomposition \eqref{nu:pert} does not apply. Even among $\nu$-analytic and $z$-independent functions, it is possible that some solutions have been overlooked. This brings us back to \eqref{w0:cons}. Aside from functions that are annihilated by the Fock projector, Eq. \eqref{w0:cons} is satisfied for even functions of the form
\begin{equation}
f(y\bar y)=f(-y\bar y)\,.    
\end{equation}
This is simply a consequence of the fact that $C_0$ itself is a function of $y\bar y$, and for even functions, we have $\pi f(y\bar y)=f(y\bar y)$. Assuming such functions solve in addition \eqref{coh}, we observe that they have to be star-product periodic. Indeed, from the vacuum equation 
\begin{equation}
    \dr_{z}\Omega+\{\Lambda_0, \Omega\}_*+\dr_x\Lambda_0=0
\end{equation}
and \eqref{Lambda0} it follows from the $\dr\rmz$-sector that
\begin{equation}\label{L:adj}
    [D, \Lambda_0]_*=\Lambda_0\,,
\end{equation}
where $D$ is the dilatation operator defined in \eqref{D}. Therefore, denoting 
\begin{equation}
    f(y\bar y):=F_*(D)\,,
\end{equation}
where $F_*(D)$ is a star-product series in $D$ and using that\footnote{Since it is not important for the following, below we take $f$ as a space-time 0-form. This explains presence of the commutator instead of the anticommutator.}
\begin{equation}
    \mathcal{D}_{\nu}f=[f, \Lambda_{0}]_*=0\,,
\end{equation}
along with \eqref{L:adj}, we have
\begin{equation}
    F_*(D)=F_*(D-1)\,.
\end{equation}
The latter implies
\begin{equation}
    F_*(D)=\sum_n a_n\exp_*(2\pi i n D)\,,
\end{equation}
where $a_n$ are arbitrary numbers. 
In the Appendix E, we demonstrate that although functions $F_*$ formally exist, the period of these is such that their symbols are not analytic in the variable $y\bar y$, which means they should be disregarded. 

\subsection{Linearizing with $\w(y^+)$}
Let us emphasize that we do not have the complete set of solutions of \eqref{coh}. In our approach, we consider only $\w(y^+)$ as the physical cohomologies for the fluctuating 1-forms. While it is generally inconsistent to restrict HS modules to submodules at the level of equations, it turns out that $\w(y^+)$ constitutes a closed sector in the broken phase, at least at the linearized level, as we will demonstrate.

The choice of the physical 1-forms restricted by $\w(y^{+})$ strongly constrains the form of the HS vertices and sets constraints on the Weyl module $\mathcal{C}$. Specifically, the solution of \eqref{lin: dL} for the 1-form master field $W$ is now organized as follows:
\begin{equation}\label{W:str}
    W=\w(y^+)+{W}'[\Omega, \mathcal{C}, C_0](z; Y)+{W}''[\w, C_0](z; Y)\,,
\end{equation}
where the first term is $z$-independent physical gauge field, representing a part of the freedom in homogeneous solutions of \eqref{lin: dL},  the second contains $z$-dependent corrections expressed in terms of the background fields and the 0-form $\mathcal{C}$, while the third term contains a contribution of the propagating 1-form $\w$. However, the latter cannot appear because the contribution containing $\w$ drops off from \eqref{lin: dL},  
\begin{equation}
    \{\w(y^+), \Lambda_0\}_*=0\quad\Rightarrow\quad {W}''[\w, C_0]=0\,. 
\end{equation}
This implies that not every structure on the right-hand side of \eqref{w:lin} is actually present. Indeed, the first term on the right comes from the third term in \eqref{W:str} upon substitution into the anticommutator of \eqref{lin:dw} followed by setting $z=0$. Since ${W}''=0$, such contribution should be equal to zero. Analogously, the first term on the right of \eqref{C:lin} should be absent. From the point of view of the HS vertices \eqref{verC}, the latter statement is a quite nontrivial one and we double-check it by a direct calculation in the Appendix C.  Additionally, Eq. \eqref{Fock:anih} implies that 
\begin{equation}
    \w*C_0-C_0*\pi(\w)=0\,.
\end{equation}
Eventually, the system \eqref{lin:unfld} takes a simpler form 
\begin{subequations}\label{slin:unfld}
    \begin{align}
       &\dr_x\w+\{\Omega, \w\}_*=\sum_n\mathcal{V}_n(\Omega, \Omega, \mathcal{C},C_0^{n-1})\,,\label{sw:lin}\\
       &\dr_x\mathcal{C}+\Omega*\mathcal{C}-\mathcal{C}*\pi(\Omega)
    =\sum_n\Upsilon_n(\Omega, \mathcal{C},
    C_0^{n})\,.\label{sC:lin}
    \end{align}
\end{subequations}
In general, omitting certain terms from the unfolded equations leads to inconsistencies. In our case, the restriction on cohomology, given by $\w(y, \bar y)=\w(y^+)$, arises from our choice of solutions to the equation \eqref{lin: dL}. This should impose a constraint on the module $\mathcal{C}$, which is yet to be determined.    

A further simplification of the system \eqref{lin:unfld} arises from the fact that $\Omega$ \eqref{Omega:vac} is no more than quadratic in $y$. Examining equation \eqref{verw} with $\Omega$ in place of $\omega$, we observe that its derivatives become zero for $n>2$. Consequently, the right-hand side of \eqref{sw:lin} also vanishes for $n>2$. In fact, it also  vanishes for $n=2$, because $\p_{\al}\p_{\gb}\Omega(y, \bar y)\wedge \p^{\al}\p^{\gb}\Omega(y', \bar y')=0$. As a result, the sum on the right of \eqref{sw:lin} contains only one  contribution corresponding to $n=1$. A similar analysis regarding the derivatives acting on $\Omega$ in \eqref{verC} leads us to conclude that only $n=1$ and $n=2$ can contribute to the right-hand side of \eqref{sC:lin}. Thus, Eqs. \eqref{lin:unfld} dramatically simplify    
\begin{subequations}\label{eqs:lin}
    \begin{align}
       &\dr_x\w+\{\Omega, \w\}_*=\mathcal{V}(\Omega, \Omega, \mathcal{C})\,,\label{w:lin-simp}\\
       &\dr_x\mathcal{C}+\Omega*\mathcal{C}-\mathcal{C}*\pi(\Omega)
    =\Upsilon(\Omega, \mathcal{C},
    C_0)+\Upsilon(\Omega, \mathcal{C},
    C_0, C_0)\,,\label{C:lin-simp}
    \end{align}
\end{subequations}
Now we only need to calculate the lowest vertices on the right-hand sides.

\subsection{Vertices: explicit form} Here we compute the manifest form of equations \eqref{eqs:lin}.

\paragraph{Vertex $\mathcal{V}(\Omega, \Omega, \mathcal{C})$} This vertex is given by a well-known expression from the central on-mass-shell theorem \cite{Vasiliev:1988sa}. Naturally, it can be obtained from \eqref{verw} for $n=1$ by substituting \eqref{Omega:vac} in place of $\omega$ with the final result being %{\color{red} Check factor}
\begin{equation}\label{V:wwC}
\mathcal{V}(\Omega, \Omega, \mathcal{C})=-\e_{\al}{}^{\dal}\wedge\e^{\al\dgb}\partial_{\dal}\partial_{\dgb}\mathcal{C}(0,\bar y)=-\frac{1}{4\rmz^2}(2i\dr\rmx^{\al\gb}\wedge\dr\rmz+\dr\rmx_{\gamma}{}^{\al}\wedge\dr\rmx^{\gb\gamma})\bar\p_{\al}\bar\p_{\gb}\mathcal{C}(0, \bar y)\,, 
\end{equation}
where $\e$ is the vierbein 1-form from \eqref{vierbein}.

Notice that the left-hand side of equation \eqref{w:lin-simp} does not match the right-hand side given by equation \eqref{V:wwC}. Specifically, while $\w$ depends on the variable $y^+$, the vertex described in \eqref{V:wwC} depends on $\bar y$. This mismatch arises because the restrictions imposed on physical fields by equation \eqref{coh} lead to a specific constraint on the Weyl module $\mathcal{C}(y, \bar y)$ through the integrability condition for \eqref{C:lin}. We will identify this constraint in the following discussion, from which it follows that
\begin{equation}\label{no HS}
    \mathcal{C}(0, \bar y)=\mathcal{C}_0+\bar y^{\al}\mathcal{C}_{\al}+\frac{1}{2}\bar y^{\al}\bar y^{\gb}\mathcal{C}_{\al\gb}\,,
\end{equation}
where the components $\mathcal{C}_0$, $\mathcal{C}_{\al}$ and $\mathcal{C}_{\al\gb}$ are $\bar y$-independent and correspond to fields of helicities 0, $-\frac{1}{2}$ and $-1$, respectively, as stated in Eq. \eqref{C:phys}. Notice that $\mathcal{C}(0, \bar y)$ cuts short at negative helicity $s=-1$. Plugging \eqref{no HS} into \eqref{V:wwC}, we obtain
\begin{equation}
    \mathcal{V}(\Omega, \Omega, \mathcal{C})=-\frac{1}{4\rmz^2}(2i\dr\rmx^{\al\gb}\wedge\dr\rmz+\dr\rmx_{\gga}{}^{\al}\wedge\dr\rmx^{\gga\gb})\mathcal{C}_{\al\gb}\,,
\end{equation}
which is $Y$-independent and captures only the helicity $s=-1$ encoded in the self-dual Maxwell tensor $\mathcal{C}_{\al\gb}$. 

\paragraph{Vertex $\Upsilon(\Omega, \mathcal{C},
    C_0)$} This vertex comes from the linearization $C\to C_0+\mathcal{C}$ of the vertex $\Upsilon(\Omega, C, C)$. The latter is known in the literature; see \cite{Vasiliev:2016xui} (specialized to the self-dual case). We reproduce the same expression in a slightly different form using \eqref{verC} with $n=2$ (see also \cite{Didenko:2024zpd}) 
\begin{equation}\label{wCC:full}
\begin{split}
    \Upsilon(\Omega, C, C)= 
    i y^{\alpha}\int_{\mathcal{D}_2^{[0]}} e^{iuv} \p_{\alpha}\gO(S_2^{[0]} y + v, \bar y) \star C(\xi_1 y - \eta_1 u,\bar y) \star C(\xi_2 y - \eta_2 u,\bar y) \\
    - iy^{\alpha}\int_{\mathcal{D}_2^{[1]}} e^{iuv} C(\xi_1 y - \eta_1 u,\bar y) \star \p_{\alpha} \gO(S_2^{[1]} y + v, \bar y) \star C(\xi_2 y - \eta_2 u,\bar y)\\
    + iy^{\alpha}\int_{\mathcal{D}_2^{[2]}} e^{iuv} C(\xi_1 y - \eta_1 u,\bar y) \star C(\xi_2 y - \eta_2 u,\bar y) \star \p_{\alpha} \gO(S_2^{[2]} y + v, \bar y)\,, 
\end{split}
\end{equation}
where the integration over $u$ and $v$ is implicitly assumed as in \eqref{verC}, while $\p$ differentiates the first argument of $\Omega$. Using \eqref{Omega:vac} and that $\mathcal{D}_2^{[0]}\sqcup\mathcal{D}_2^{[2]}=\mathcal{D}_2^{[1]}$, \eqref{domain}, the integration over $u$ and $v$ can be completed. Connection $\gO$ consists of linear and bilinear in $y$ parts. Direct calculation shows that the bilinear part contribution vanishes, which is also due to the structure of $S_n^{[k]}$; see \eqref{Snk}. Eventually, Eq. \eqref{wCC:full} can be shuffled into the following form:  %{\color{red} Check factor} 
\begin{align} 
    &\Upsilon(\Omega, C, C)=\ff{1}{2}\e^{\al\dgb}y_{\al}
\int_{\mathcal{D}_{2}^{[0]}}[[\bar y_{\dgb}, C(\xi_1 y, \bar
y)]_{\star}, C(\xi_2 y, \bar
y)]_{\star}=\nn\\
&=i\e^{\al\dgb}y_{\al}
\int_{\mathcal{D}_{2}^{[0]}}[\bar\p_{\dgb}C(\xi_1 y, \bar y),
C(\xi_2 y, \bar y)]_{\star}\,,
\end{align}
It is straightforward to linearize the above expression about \eqref{C0} with the result being %{\color{red} Check}
\begin{align}\label{Ups:wCC0}
&\Upsilon(\Omega, \mathcal{C}, C_0):=\Upsilon(\Omega, C, C)\Big|_{C=C_0+\mathcal{C}}=\\
&=i\nu\rmz\,\e^{\al\gb}y_{\al}\int_{0}^{1}d\tau\tau e^{\tau y\bar y}((1-\tau)y_{\gb}-\bar\p_{\gb})\Big(
\mathcal{C}((1-\tau)y, \bar y-i\tau y)-\mathcal{C}((1-\tau)y, \bar
y+i\tau y)\Big)\,,\nn
\end{align}
where the integration over $\mathcal{D}_2^{[0]}$ has been reduced to the one-dimensional over $\tau$ using 
\begin{equation}
    \int_{\mathcal{D}_2^{[0]}}f(\xi_1, \xi_2)=\int_{0}^{1}d\tau (1-\tau)f(\tau, 1-\tau)\,.
\end{equation}
Recall also that $\e^{\al\gb}:=\e^{\al\dgb}\gd_{\dgb}{}^{\gb}$.   

%For calculations it is convenient to split the connection into parts by powers of $y$
%\begin{equation}
%    \gO^{lin} = -\frac{i}{2}e^{\mu\dot %\nu}y_{\mu}\bar y_{\dot \nu},\ \gO^{quad} = %-\frac{i}{4}\omega^{\mu\nu}y_{\mu}y_{\nu}
%\end{equation}
%Obviously we have
%\begin{equation}
%    \Upsilon(\gO, \mathcal{C}, C_0) = %\Upsilon(\gO^{lin}, \mathcal{C}, C_0) + %\Upsilon(\gO^{quad}, \mathcal{C}, C_0)
%\end{equation}
%Now one can make sure that
%\begin{equation}
%    \Upsilon(\gO^{quad}, \mathcal{C}, C_0) = 0
%\end{equation}
%which is due to the two features: 
%And the vertex is determined by the linear part %only. Direct calculation gives
%\begin{equation}
%    \Upsilon(\gO, \mathcal{C}, C_0) = -2i\rmz C_0 %\cdot e^{\mu\dot{\nu}}y_{\mu}\bar %\partial_{\dot{\nu}} \int_0^1 d\tau\ (1 - \tau) %[T(\tau w, \bar w - i[1 - \tau]w) - T(\tau w, \bar %w + i[1 - \tau]w)]
%\end{equation}

\paragraph{Vertex $\Upsilon(\Omega, \mathcal{C},
    C_0, C_0)$}
Similarly to the previous case, this vertex follows from the linearization of $\Upsilon(\Omega, C, C, C)$. However, the vertex is equal to zero as we show now. Indeed, from \eqref{verC} it follows that the vertex contains $(y^{\al}\p_{\al})^2\Omega(S_3^{[k]}y+u, \bar
y)$, which is, thanks to the fact that $\Omega$ is bilinear in
$y$, leads to the following $u$ and $S$ independent piece:
\be
\hat\Omega:=(y^{\al}\p_{\al})^2\Omega(S_3^{[k]}y+u, \bar
y)=\ff{1}{4\rmz}y^{\al}y^{\gb}\dr\rmx_{\al\gb}\,.
\ee
This allows us to complete the integration over $u$ and $v$ in \eqref{verC} and gives
us
\begin{align}\label{contr:C3}
&\Upsilon(\Omega, C, C,
C)=(-i)^2\hat\Omega\cdot\Big(\int_{\mathcal{D}_3^{[0]}}C(\xi_1y, \bar
y)\star C(\xi_2y, \bar y)\star C(\xi_3y, \bar
y)-\\
-&\int_{\mathcal{D}_3^{[1]}}C(\xi_1y, \bar y)\star C(\xi_2y, \bar
y)\star C(\xi_3y, \bar y)+\int_{\mathcal{D}_3^{[2]}}C(\xi_1y, \bar
y)\star C(\xi_2y, \bar y)\star C(\xi_3y, \bar y)-\nn\\
-&\int_{\mathcal{D}_3^{[3]}}C(\xi_1y, \bar y)\star C(\xi_2y, \bar
y)\star C(\xi_3y, \bar y)\Big)\,.\nn
\end{align}
Now, the property of the integration domain \eqref{domain}, which in this case reads $\mathcal{D}_3^{[0]}\sqcup\mathcal{D}_3^{[2]}=\mathcal{D}_3^{[1]}\sqcup\mathcal{D}_3^{[3]}$
makes \eqref{contr:C3} identically vanish. So, we conclude that 
\begin{equation}
    \Upsilon(\Omega, C, C,
C)=0\quad\Rightarrow\quad \Upsilon(\Omega, \mathcal{C},
    C_0, C_0)=0\,.
\end{equation}

Let us summarize our results on the free field equations obtained at this point. As follows from the above computations, the linear system \eqref{eqs:lin} is further reduced. In particular, it lacks $O(\nu^2)$ contribution represented by the last term in \eqref{C:lin}. The $O(\nu)$ contribution is given by \eqref{Ups:wCC0}, while the gauge sector \eqref{w:lin} contains no $\nu$-dependent terms at all:
\begin{subequations}\label{lin}
    \begin{align}
       &\dr_x\w+\{\Omega, \w\}_*=-\e_{\al}{}^{\dal}\wedge\e^{\al\dgb}\partial_{\dal}\partial_{\dgb}\mathcal{C}(0,\bar y)=-\e_{\al}{}^{\dal}\wedge\e^{\al\dgb}\mathcal{C}_{\al\gb}\,,\label{dw:lin}\\
       &\dr_x\mathcal{C}+\Omega*\mathcal{C}-\mathcal{C}*\pi(\Omega)
    =\Upsilon(\Omega, \mathcal{C},
    C_0)\,.\label{dC:lin}
    \end{align}
\end{subequations}
The source for the gauge fields on the right side of equation \eqref{dw:lin} is zero for all helicities except for $s=-1$. As a result, for non-positive helicities we are left with only one gauge field of $s=-1$, along with matter fields of spins $s=0$ and chiral $s=-\frac{1}{2}$, \eqref{no HS}.

\paragraph{Consistency}
Consistency of the system \eqref{lin} is not guaranteed despite the original equations \eqref{Eqs} from which it comes being consistent. The reason is a restriction set on the physical field \eqref{w:phys}, which results in certain vertex structures vanishing. Therefore, we need to check if \eqref{lin} respects integrability $\dr_x^2=0$. 

Applying $\dr_x$ to \eqref{dw:lin} introduces no new constraints, confirming its consistency. The proof is relatively straightforward; it involves substituting $\dr_x\mathcal{C}_{\al\gb}$ from \eqref{dC:lin}, and notably, no contributions from $\Upsilon(\Omega, \mathcal{C}, C_0)$ \eqref{Ups:wCC0} appear during this process.
However, the consistency of \eqref{dC:lin} is not immediately obvious. In fact, \eqref{dC:lin} will only be consistent if the condition \eqref{no HS} is satisfied. Since the proof is quite technical, we have included it in the Appendix D.

\section{Dual formulation}\label{sec:dual}
Klebanov and Polyakov conjectured holographic duality between Vasiliev's HS theory and conformal $O(N)$ models. The basis of the conjectured duality is the fundamental result of Flato and Fronsdal, who showed that an infinite tower of the $4d$ massless fields can be viewed as tensor products of $3d$ singletons \cite{Flato:1978qz}. The map between the AdS and conformal fields is particularly simply realized using the unfolded approach at the level of equations of motion, as demonstrated in \cite{Vasiliev:2012vf}. The natural question we ask here is how the Flato-Fronsdal result modifies in the symmetry-broken phase. To this end, let us first recall the derivation in the unbroken phase with $\nu=0$ following \cite{Vasiliev:2012vf}. 

\subsection{Symmetry unbroken case: $\nu=0$}

Since field degrees of freedom are collected in the 0-form sector, we consider Eq. \eqref{dC:lin} with $\nu=0$
\begin{equation}
    \dr_x\mathcal{C}_{\nu=0}+\Omega*\mathcal{C}_{\nu=0}-\mathcal{C}_{\nu=0}*\pi(\Omega)=0\,,
\end{equation}
where by $\mathcal{C}_{\nu=0}$ we denoted the standard twisted-adjoint module corresponding to \eqref{dC:lin} with vanishing right-hand side. Substituting \eqref{Omega:vac} and using \eqref{star:gen}, one arrives at %{\color{red} check below}
\begin{subequations}\label{tw-ad:3+1}
    \begin{align}
        &\dr_{\rmx}\mathcal{C}_{\nu=0}+\frac{i}{2\rmz}\dr\rmx^{\al\gb}(y-\bar\p)_{\al}(\bar y+\partial)_{\gb}\,\mathcal{C}_{\nu=0}=0\,,\\
        &\frac{\partial}{\partial \rmz}\mathcal{C}_{\nu=0}+\frac{1}{2\rmz}(y\bar y-\partial\bar\partial)\,\mathcal{C}_{\nu=0}=0\,,
    \end{align}
\end{subequations}
where the first equation of \eqref{tw-ad:3+1} describes the evolution of the Weyl module along a three-dimensional AdS slice, while the second -- its dependence on the radial direction. The Flato-Fronsdal correspondence can be approached by introducing the intertwining of the Weyl module with the current module $T$ as follows:
\begin{equation}\label{T:def}
    \mathcal{C}=\rmz\,e^{y\bar y}\,T(w, \bar w|\vec\rmx, \rmz)\,,\qquad w=\sqrt{\rmz}y\,,\quad \bar w=\sqrt{\rmz}\bar y\,.
\end{equation}
We will use the same map for arbitrary $\nu$ because the background geometry is AdS. Notice the presence of the vacuum \eqref{C0} in the definition \eqref{T:def}. In terms of the current module $T$, the system \eqref{tw-ad:3+1} reduces to 
\begin{subequations}\label{T:cons}
    \begin{align}
        &\dr_{\rmx}T_{\nu=0}-\frac{i}{2}\dr\rmx^{\al\gb}\p_{\al}\bar\p_{\gb}T_{\nu=0}=0\,,\label{Tx:consv}\\
        &\frac{\p}{\p\rmz}T_{\nu=0}-\frac{1}{2}\p\bar\p\,T_{\nu=0}=0\,.\label{Tz:consv}
    \end{align}
\end{subequations}
Here $\p$ and $\bar \p$ denote differentiation with respect to the spinorial argument in both \eqref{tw-ad:3+1} and \eqref{T:cons}, e.g., $\p_{\al}\mathcal{C}:=\frac{\p}{\p y^{\al}}\mathcal{C}$, $\bar \p_{\al}{T}:=\frac{\p}{\p \bar w^{\al}}{T}$, while $\frac{\p}{\p\rmz}$ differentiates the manifest dependence on $\rmz$; in particular, $\frac{\p}{\p\rmz} w=0$.  Notice that manifest dependence on $\rmz$ completely decouples from \eqref{T:cons}.  Equation \eqref{Tx:consv} describes a set of conserved $3d$ currents
\begin{equation}\label{currents}
    \frac{\p}{\p\rmx^{\al_1\al_2}}j^{\al_1\dots\al_{s}}=0\,,\qquad  \frac{\p}{\p\rmx^{\al_1\al_2}}\bar j^{\al_1\dots\al_{s}}=0\,,
\end{equation}
which are associated with coefficients in the decomposition of $j(w):=T(w, 0)$ and $\bar j(\bar w):=T(0, \bar w)$. It also describes a primary of conformal dimension $\Delta=1$, $j_0=T(0,0)$, and a primary of conformal dimension $\Delta=2$, $\tilde{j}_0:=\p\bar \p T(w,\bar w)|_{w=\bar w=0}$. These appear as $H^0(\gs^-)$-cohomologies, where 
\begin{equation}\label{sigma}
    \gs^-=\dr\rmx^{\al\gb}\frac{\p}{\p w^{\al}}\frac{\p}{\p\bar w^{\gb}}\,.
\end{equation}

The Flato-Fronsdal decomposition then arises as
\begin{equation}\label{FF}
    T_{\nu=0}=\sum_{i}C_+^i(w^+)C_-^i(w^-)\,,\qquad w^{\pm}_{\al}=w_\al\pm i\bar w_\al\,,
\end{equation}
where $C^i_{\pm}$ are subject to the following condition:
\begin{equation}\label{scalareq}
    \dr_{\rmx}C_{\pm}^{i}\pm\frac{1}{2}\dr\rmx^{\al\gb}\p_\al\p_\gb C^i_{\pm}=0\,.
\end{equation}
It is straightforward to verify that \eqref{Tx:consv} is satisfied for \eqref{FF}, provided \eqref{scalareq} is imposed. The latter just describes free massless scalars on the $3d$ Minkowski background 
\begin{equation}
    \Box\varphi^i_{\pm}=0\,,\qquad \varphi^i_{\pm}:=C^i_\pm(0|x)\,.
\end{equation}

\subsection{Symmetry broken case: $\nu\neq 0$} Now we are interested in how Eqs. \eqref{T:cons} are modified in the broken phase with $\nu\neq 0$. To see this, we use the field-current correspondence \eqref{T:def} and substitute it into \eqref{dC:lin} using \eqref{Ups:wCC0}. After simple algebra, we arrive at the following final result: %{\color{red} check coefficients} 
\begin{subequations}\label{T:nonconsv}
    \begin{align}
        &\dr_{\rmx}T-\frac{i}{2}\dr\rmx^{\al\gb}\p_{\al}\bar\p_{\gb}T=\ff{i\nu}{2}\dr\rmx^{\al\gb}w_{\al}\bar\p_{\gb}\int_{0}^{1}d\tau(1-\tau)
(T^--T^+)\,,\label{Tx:nonconsv}\\
        &\frac{\p}{\p\rmz}T-\frac{1}{2}\p\bar\p\,T=-\frac{\nu}{2} w^{\al}\bar\p_{\al}\int_{0}^{1}d\tau(1-\tau)
(T^--T^+)\,,\label{Tz:nonconsv}
    \end{align}
\end{subequations}
where 
\begin{equation}
   T^{\pm}=T(\tau w, \bar w\pm i(1-\tau)w)\,.
\end{equation}
Notice that the factor $\exp{y\bar y}$ in \eqref{T:def}, as well as manifest dependence on $\rmz$, decouples from \eqref{T:nonconsv}. Thus, one indeed arrives at the dual formulation of the on-shell $4d$ system via the $3d$ off-shell given by \eqref{T:nonconsv}. The crucial question, of course, is its consistency. As we stated earlier, the original equations \eqref{dC:lin} are generally inconsistent, unless the Weyl module $\mathcal{C}$ is constrained by \eqref{no HS}. The latter condition imposes a constraint on $T$ of the form 
\begin{equation}\label{no hc}
    T(0, \bar w)=T_0+\bar w^{\al}T_{\al}+\frac{1}{2}\bar w^{\al}\bar w^{\gb}T_{\al\gb}\,.
\end{equation}
In the Appendix D, we check that \eqref{no hc} indeed follows from $\dr_\rmx^2=\{\dr_{\rmx}, \dr_{\rmz}\}=0$. Constraint \eqref{no hc}, which can be reformulated as
\begin{equation}
    \left(\frac{\p}{\p \bar w}\right)^3T(0, \bar w)=0\,,
\end{equation}
generates a secondary constraint from 
\begin{equation}
    \dr_x \left(\frac{\p}{\p \bar w}\right)^3T(0, \bar w)=\left(\frac{\p}{\p \bar w}\right)^3\dr_x T(0, \bar w)=0\,,
\end{equation}
and so forth. It is not difficult to identify them all by noting that at $w=0$ Eq. \eqref{Tx:nonconsv} amounts to conservation of chiral currents encoded in
\begin{equation}
    \dr_\rmx T(0, \bar w)-\frac{i}{2}\dr\rmx^{\al\gb}\p_{\al}\bar\p_{\gb}T(w, \bar w)|_{w=0}=0\,.
\end{equation}
Constraint \eqref{no hc} then implies that the helicity-$s$ current modules $T_s$ have to vanish for $s<-1$, 
\begin{equation}\label{no s>1}
    T_{s<-1}=0\,,
\end{equation}
where
\begin{equation}
    \left(w^{\al}\p_{\al}-\bar w^{\al}\bar\p_{\al}\right)T_s=2sT_s\,.
\end{equation}
 Notice that the whole module $T$ splits into a direct sum: 
 \begin{equation}\label{T:decomp}
     T=\bigoplus_s T_s\,,\qquad s=0, \pm \frac{1}{2}, \pm 1\,\dots\,,
 \end{equation}
 while \eqref{no hc} factors out a set of submodules $T_{s<-1}$, keeping the rest $T_{s\geq -1}$ alive.  

\paragraph{Primaries} Equation \eqref{Tx:nonconsv} is supposed to describe the broken phase of the current conservation condition. In the unbroken case, the conserved currents \eqref{currents} arise as the following consequence of \eqref{Tx:consv}:
\begin{equation}
    \frac{\p}{\p\rmx_{\al\gb}}\frac{\p}{\p w^{\al}}\frac{\p}{\p w^{\gb}}j(w)=\frac{\p}{\p\rmx_{\al\gb}}\frac{\p}{\p\bar w^{\al}}\frac{\p}{\p\bar w^{\gb}}\bar j(\bar w)=0\,,
\end{equation}
where the current primaries are associated with 0-form cohomologies of \eqref{sigma} 
\begin{equation}
    j(w)=T(w, 0)\,,\qquad \bar j(\bar w)=T(0, \bar w)\,.
\end{equation}
When $\nu\neq 0$, Eq. \eqref{Tx:consv} is replaced with \eqref{Tx:nonconsv}, which we can write down as 
\be\label{x-eq:sigma}
\dr_{\rmx}T(w, \bar w)-\ff{i}{2}\gs_{\nu} T(w, \bar w)=0\,,
\ee
where
\be\label{sigma:nu}
\gs_{\nu} T(w, \bar w):=\dr\rmx^{\al\gb}\left(\p_{\al}\bar \p_{\gb}T(w,
\bar w)+
\nu\,w_{\al}\bar\p_{\gb}\int_{0}^{1}d\tau(1-\tau)
(T^--T^+)\right)\,.
\ee
Primary fields cannot be expressed via derivatives of other fields. Thus, those belonging to the kernel of $\gs_{\nu}$ are necessarily primaries. We identify them by solving for primary components $T^{pr}$
\begin{equation}\label{coh:sigma}
    \gs_{\nu}T^{pr}(w, \bar w)=0\,.
\end{equation}  
To do so, we note that \eqref{coh:sigma} is equivalent to
\begin{equation}\label{eq:psi}
    \bar\p_{\gb}\psi_{\al}+\bar\p_{\al}\psi_{\gb}=0\,,
\end{equation}
where
\be\label{psi}
\psi_{\al}=\p_{\al}T^{pr}+\nu\,w_{\al}\int_{0}^{1}d\tau(1-\tau)
(T^{pr -}-T^{pr +})\,.
\ee
The general solution of \eqref{eq:psi} is
\be
\psi_{\al}=\phi_{1\al}(w)+\bar w_{\al}\phi_{2}(w)\,,
\ee
where $\phi_{1,2}$ are arbitrary functions. 
Contracting \eqref{psi} with $w^{\al}$, we have
\be
w^{\al}\p_{\al}T^{pr}=w^{\al}\phi_{1\al}(w)-w\bar w\,\phi_{2}(w)\,.
\ee
Therefore, $T^{pr}$ itself should be of the following form
\be
T^{pr}=j(w)+w\bar w\, f(w)+\bar j(\bar w)\,,
\ee
where $j$ and $f$ are expressed via $\phi_1$ and $\phi_2$, and $\bar j$ is arbitrary. 
Substituting it back to \eqref{coh:sigma}, it is easy to see that the equation is only satisfied for the following functions:
\be\label{deform:coh}
j(w)\,,\qquad w\bar w\,\tilde{j}_0\,,\qquad \bar j_{\al}\bar w^{\al}\,,
\ee
where $\tilde{j}_0$ and $\bar j_{\al}$ do not depend on $w$, $\bar w$. The list of these primaries may not be complete. Specifically, $\gs_{\nu}$ does not respect grading as it  mixes various components of the module $T$. As a result, if $\gs_{\nu}T\neq 0$, then a combination of several components of $T$ is expressed via a derivative of a single one. Therefore, it is a matter of definition which field from the set generated by $\gs_{\nu}$ should be called primary. In particular, the field associated with the quadratic combination of $\bar w$ (the helicity $s=-1$ antichiral current), 
\begin{equation}\label{s=1: acc}
  \bar j_1(\bar w)=\frac{1}{2}\bar j_{\al\gb}\bar w^{\al}\bar w^{\gb}  
\end{equation}
can be taken as a primary. Indeed, using \eqref{sigma:nu} we find that 
\begin{equation}
    \gs_{\nu}\bar j_1(\bar w)=\frac{2i\nu}{3}\dr\rmx_{\al}{}^{\gga}\bar j_{\gga\gb}w^{\al}w^{\gb}
\end{equation}
contributes to the following sector of \eqref{Tx:nonconsv}:
\begin{equation}\label{s=1: acc pr}
    \dr_{\rmx}j_{\al(2)}-\frac{i}{2}\dr\rmx^{\gb(2)}T_{\gb\al(2), \gb}+\frac{2\nu}{3}\dr\rmx_{\al}{}^{\gb}\bar j_{\gb\al}=0\,,
\end{equation}
where we use the notation 
\begin{equation}\label{T:comp}
    T(w, \bar w):=\sum_{m,n}\frac{1}{m!n!}T_{\al(m), \gb(n)}(w^{\al})^n(\bar w^{\gb})^n\,,\quad j_{\al(m)}:=T_{\al(m),\gb(0)}\,,\quad \bar j_{\gb(n)}:=T_{\al(0),\gb(n)}\,.
\end{equation}
From \eqref{s=1: acc pr} we see that the field $T_{\gb\al(2), \gb}$ is expressed via the derivative of a primary $j_{\al(2)}$ and the field $\bar j_{\al(2)}$. The latter can be chosen as a primary. Given that the module $T$ decomposes into a direct sum \eqref{T:decomp} and in view of \eqref{no s>1}, it is not hard to make sure that the full list of primaries is 
\begin{equation}
    j(w)\,,\qquad w\bar w\,\tilde{j}_0\,,\qquad \bar j_{\al}\bar w^{\al}\,,\qquad \bar j_{\al\gb}\bar w^{\al}\bar w^{\gb}\,.
\end{equation}

\paragraph{Component form} Using the definition \eqref{T:comp}, let us rewrite \eqref{Tx:nonconsv}, \eqref{Tz:nonconsv} in terms of components of $T$. We can merge the two equations into one via the vierbein \eqref{vierbein} 
\begin{equation}
    \dr_x T = -i\rmz \e^{\alpha\beta}(\p_{\alpha}\bar\p_{\beta}T + \nu w_{\alpha}\bar\p_{\beta}J[T])\,,
\end{equation}
where we denote
\begin{equation}
    J[T] : = \int_0^1 d\tau (1 - \tau) (T^- - T^+)\,.
\end{equation}
This form is convenient, because it contains both symmetric and anti-symmetric parts of the spinor components on the right-hand side. Substitution of \eqref{T:comp} gives us
\begin{equation}\label{Teq:comp}
    \dr_{x}T_{\al(m), \gb(n)} = -i\rmz \e^{\gamma\rho}\left(T_{\gamma\alpha(m),\rho\beta(n)} + 4i\nu \epsilon_{\alpha\gamma} \sum_{k = 1}^{\lfloor \frac{m}{2} \rfloor} \frac{k(-)^{k}}{m + 1}T_{\alpha(m - 2k),\alpha(2k - 1)\beta(n)\rho}  \right)\,.
\end{equation}
%\begin{equation}\label{Teq:comp}
%    \dr_{\rmx}T_{\al(m), \gb(n)}-\frac{i}{2}\dr\rmx^{\gga(2)}T_{\gga\al(m), %\gga\gb(n)}=-2\nu\dr\rmx^{\gga}{}_{\al}\sum_{k=1}^{[m/2]}\frac{(-)^k k}
%{m+1}T_{\al(m-2k), \al(2k-1)\gb(n)\gga}\,.
%\end{equation}
%In obtaining this result we have not taken into account the consistency condition \eqref{no s>1}, the component form of which is 
%\begin{equation}
%    T_{\al(m+2s), \gb(m)}=0\quad\text{for all}\quad s<-1\,,\quad m+2s\geq 0\,.
%\end{equation}
In obtaining this result we have not taken into account the consistency condition \eqref{no hc}, the component form of which is 
\begin{equation}
    \bar j_{\gb(n)}:=T_{\al(0), \gb(n)}=0\quad\text{for all}\quad n > 2\,.
\end{equation}
This condition sets to zero all the components $T_{\al(m), \gb(n)}$ with $n - m > 2$. To see this, one starts with the \eqref{Teq:comp} for the $m = 0, 1$ cases and finds $T_{\alpha, \beta(n + 1)} = 0\,,\quad T_{\alpha(2), \beta(n + 2)} = 0 \quad\text{for all}\quad n > 2$. The general claim is proven by induction (see Appendix D)
\begin{equation}
    T_{\al(m+2s), \gb(m)}=0\quad\text{for all}\quad s<-1\,,\quad m+2s\geq 0\,.
\end{equation}
For the right-hand side of \eqref{Teq:comp} the latter implies that the sum is nonzero only if
\begin{equation}\label{ineq}
    m-4k-n\geq-2\quad\Rightarrow\quad k\leq\frac{m+2-n}{4}\,.
\end{equation}
It is convenient to further rewrite \eqref{Teq:comp} in terms of the helicity-$s$ modules $T^{s}$,  \eqref{T:decomp}, which means that one simply has to replace indices as follows $m\to m+2s$ and $n\to m$. For the 3D part we have 
\begin{equation}\label{Tx:comp}
    \dr_{\rmx}T^s_{\al(m+2s), \gb(m)} - \frac{i}{2}\dr\rmx^{\gamma\gamma}T^{s}_{\gamma\alpha(m+2s),\gamma\beta(m)} = 2\nu\dr\rmx^{\gamma}{}_{\alpha} \sum_{k = 1}^{\lfloor \frac{s + 1}{2} \rfloor} \frac{k(-)^{k}}{m + 2s + 1}T^{s - 2k}_{\alpha(m + 2s - 2k),\alpha(2k - 1)\beta(m)\gamma}
\end{equation}
%\begin{equation}\label{Tx:comp}
%    \dr_{\rmx}T^{s}_{\al(m+2s), \gb(m)}-\frac{i}
%{2}\dr\rmx^{\gga(2)}T^{s}_{\gga\al(m+2s), \gga\gb(m)}=-2\nu\dr\rmx_{\al}
%{}^{\gga}\sum_{k=1}^{\left[\frac{s+1}{2}\right]}\frac{(-)^kk}{m+2s+1} 
%T^{s-2k}_{\al(m+2s-2k), \al(2k-1)\gb(m)\gga}\,,
%\end{equation}
where the constraint \eqref{ineq} has been taken into account. Let us summarize the main features of Eq. \eqref{Tx:comp}.
\begin{itemize}
    \item Equation \eqref{Tx:comp} acquires the following schematic form: 
\begin{equation}\label{schem}
    DT^s\sim\nu\,\dr\rmx\sum_{s'=-1}^{s-2}T^{s'}\,,\qquad s\geq 1\,,
\end{equation}
where by $D$ we sloppily denoted the action on the module $T^s$ manifested on the left of \eqref{Tx:comp}. In particular, when $\nu=0$ the equation $DT^s=0$ is a standard current-conserving condition \eqref{Tx:consv} for helicity-$s$ field. In the case $\nu\neq 0$ the left-hand side of \eqref{schem} is generally nonzero unless $s-2<-1$, i.e.,
\begin{equation}
    \dr_{\rmx}T^{s}_{\al(m+2s), \gb(m)}-\frac{i}{2}\dr\rmx^{\gga(2)}T^{s}_{\gga\al(m+2s), \gga\gb(m)}=0\quad\text{for}\quad s=-1, 0, \pm\frac{1}{2}\,,\quad\forall\nu\neq 0 
\end{equation}
\item If $s\geq 1$ the right-hand side is nonzero, while the $s$-current $j^s$ no longer conserves in general. Indeed, from \eqref{Tx:comp} we have (recall, $j_{\al(2s)}:=T_{\al(2s), \gb(0)}$)
%\begin{equation}
%    \dr_\rmx j^s_{\al(2s)}-\frac{i}
%{2}\dr\rmx^{\gga(2)}T^s_{\gga\al(2s),\gga}=-2\nu\,
%\dr\rmx_{\al}{}^{\gga}\sum_{k=1}^{\left[\frac{s+1}{2}\right]}\frac{(-)^kk}{2s+1}
%T^{s-2k}_{\al(2s-2k), \al(2k-1)\gga}\,.
%\end{equation}
\begin{equation}
    \dr_{\rmx}j_{\al(2s)} - \frac{i}{2}\dr\rmx^{\gamma(2)}T^{s}_{\gamma\alpha(2s),\gamma} = 2\nu\dr\rmx^{\gamma}{}_{\alpha} \sum_{k = 1}^{\lfloor \frac{s + 1}{2} \rfloor} \frac{k(-)^{k}}{2s + 1}T^{s - 2k}_{\alpha(2s - 2k),\alpha(2k - 1)\gamma}
\end{equation}
From here we find after some combinatorial algebra 
\begin{equation}\label{div: j}
   \frac{\p}{\p \rmx^{\gb(2)}}j^{\gb(2)}{}_{\al(2s-2)}= -\nu\sum_{k=1}^{\lfloor\frac{s+1}{2}\rfloor}\frac{(-)^k(s - k)(2k)!}{s(2s - 1)}T^{s-2k}_{\gb\al(2s-2k-1), \al(2k-1)\gamma}\epsilon^{\gb\gamma}\,.
\end{equation}
The right-hand side above contains traces of the current modules $T^{s'}$ that depend on the radial coordinate $\rmz$ via \eqref{Tz:nonconsv}. This means that these traces are not defined via other currents in accordance with the fact that the global higher-spin symmetry is broken rather than deformed. Therefore, the $s\geq 1$ currents are generally not conserved with the exception of $s=1$ for which it follows from \eqref{div: j}
\begin{equation}
    \frac{\p}{\p \rmx^{\al\gb}}j^{\al\gb}_{s=1}=0\,.
\end{equation}
\end{itemize}

\section{Concluding notes}\label{sec:conc}
In this paper, we have addressed the problem of the higher-spin symmetry breaking in a toy model example: the holomorphic sector of the 4D higher-spin theory, also known as self-dual or chiral. The standard higher-spin vacuum is invariant under AdS isometries and, more broadly, under global higher-spin symmetry, where the AdS algebra is a maximal finite-dimensional subalgebra. We mildly break this symmetry by modifying the vacuum, introducing a nonzero scalar expectation that is only nonzero in the AdS radial direction associated with the Poincar\'{e} coordinate $\rmz$. This scalar profile represents a mixture of two conformal branches, characterized by two arbitrary parameters. For simplicity, we focus on a one-parameter vacuum where the scalar has a conformal dimension of $\Delta=1$, setting the parameter associated with the branch $\Delta=2$ to zero. The remaining parameter $\nu$ represents a scale of symmetry breaking in our model. The leftover symmetry is described by the Poincar\'{e} algebra in three dimensions. In other words, the symmetry of a 3D AdS slice $\rmz=const$ is unbroken. 

As a classical solution, the obtained vacuum is perfectly regular (see Eq. \eqref{C0}). Its unfolded realization in the star-product algebra introduces a Fock projector that simplifies the analysis of nonlinear vertices by reducing dependence on the generating oscillators. The Fock projector results in a substantial simplification of the linearized dynamics on the symmetry-broken vacuum that requires accounting for all-order in $\nu$ contributions. As a side remark, it is of interest that Fock projectors frequently appear in higher-spin theory, from black holes \cite{Didenko:2009td, Iazeolla:2011cb, Kraus:2012uf, Didenko:2021vdb} and classical solutions \cite{Iazeolla:2007wt, Aros:2017ror} to holography \cite{Vasiliev:2012vf, Didenko:2012vh, Didenko:2012tv, Didenko:2017lsn, David:2020fea, Lysov:2022zlw}.

The analysis of the free field dynamics has been carried out using the holomorphic generating system of \cite{Didenko:2022qga}, which gives us access to all-order interacting vertices in their maximally local form \cite{Didenko:2024zpd}. Typically of the Vasiliev construction, free fields are described by 1-form gauge fields and 0-form generalized Weyl tensors. The gauge fields emerge as cohomologies of the operator that depends on the vacuum. For the standard vacuum, this operator is simply the de Rham operator $\dr_z$, while the gauge fields appear as $z$-independent functions. In our case with the symmetry-broken vacuum, the operator takes a $\nu$-dependent form \eqref{D:oper}. Finding cohomologies of this operator is challenging. We have focused our analysis on the $z$-independent cohomologies, which have been successfully identified. However, since this set is not complete, consistency constraints arise that limit the 0-form module accordingly. Specifically, among the Weyl 0-forms of all helicities, only those with helicities $s\geq -1$ are permitted by consistency, while those with $s<-1$ must be excluded. 

Let us stress once again that our consideration of free field dynamics is limited to excitations with $s\geq -1$. This analysis reveals several important features:

\begin{itemize}
    \item The dependence on $\nu$ completely disappears in the gauge field sector, while only linear in $\nu$ terms contribute to the Weyl sector of 0-forms. The cancellation of higher-order in $\nu$ terms occurs due to our selection of gauge fields, the characteristics of the vacuum Fock projector, and a unique structure of the holomorphic vertices, such as a specific property of the vertex integration domains \eqref{domain} that has not been previously recognized. 
   
    \item Among the infinitely many Weyl tensors with helicities $s\geq -1$, only the case $s=-1$ sources the gauge field, implying that the higher-spin gauge sector decouples for all non-negative helicities. For $s=-1$, there is the self-dual Maxwell source. 

    \item The Weyl sector also disentangles from the gauge fields. Its dual form corresponds to certain current equations, but unlike the standard Flato and Fronsdal case, these currents generally do not conserve. Furthermore, the nonconservation of the helicity $s$ current involves a contribution from all the allowed helicities $s'$ with $s'\leq s-2$. In contrast, lower spins $s=\pm 1,\ \pm 1/2,\ 0$ exhibit different behavior: their equations are either uneffected by symmetry breaking or, as in the case of $s=1$, the corresponding current remains conserved.    
\end{itemize}

Let us discuss some aspects of the linearized analysis of the broken phase that have not been fully addressed in our investigation. Firstly, since we do not have the complete set of cohomologies of the vacuum operator \eqref{D:oper}, we cannot approach fields with helicities  $s<-1$. We do not anticipate these fields to decouple, nor do we expect the corresponding currents to become independent of the gauge fields. This kind of analysis is crucial for gaining a comprehensive understanding of free field dynamics around the vacuum we are considering. However, it is likely to be more complex, as we do not expect that higher-order terms in $\nu$ will be absent in this case.

Another interesting area for future research is to explore the effects of the vacuum parameter $\nu_2$  in \eqref{C:vac}, which we set to zero for simplicity. We hope to report on these problems elsewhere.

\section*{Acknowledgments}
We would like to thank A. Korybut, N. Misuna for fruitful discussions. We also thank A. Korybut, D. Ponomarev, M. Povarnin, A. Tarusov, K. Ushakov, and M.A. Vasiliev for their comments on the draft and valuable suggestions. We are grateful for the financial support from the Foundation for the Advancement of Theoretical Physics and Mathematics “BASIS.”

%Since conformal symmetry is broken, the dilatation operator $D$ is no longer a proper object for distinguishing primaries as its eigenstates. 

    % основной текст -- от абстракта до acknowledgments
    
\newcounter{appendix}
\setcounter{appendix}{1}
\renewcommand{\theequation}{\Alph{appendix}.\arabic{equation}}
\addtocounter{section}{1} \setcounter{equation}{0}
 \renewcommand{\thesection}{\Alph{appendix}.}

\addcontentsline{toc}{section}{\,\,\,\,\,\,\,A. Deriving the vacuum}

\section*{A. Deriving vacuum}
Here we derive the vacuum solution \eqref{W:vac}, \eqref{C:vac}, which solves equations \eqref{Eqs}. For convenience, let us reproduce again the generating equations we need to solve 
\begin{subequations}\label{app:eqs}
    \begin{align}
        &\dr_x W + W * W = 0\label{eq:dWapp}\,, \\
        &\dr_z W + \{W,\Lambda\}_* + \dr_x\Lambda = 0\label{eq:dLapp}\,, \\
        &\dr_x C + \left(W(z', y)*_y C-C*_y W(z', -y)\right)\Big|_{z' = -y} = 0\,,\label{eq:dCapp}
    \end{align}
\end{subequations}
where
\begin{equation}
    \Lambda = \dr z^{\alpha}z_{\alpha}\int_0^1 d\tau\tau C(-\tau z, \bar y) e^{i\tau zy}\,.
\end{equation}
We start out with an ansatz that assumes the vacuum connection $W$ is exactly the same as in the AdS space \eqref{Omega:vac}, while the field $C$ is nonzero and contains scalar degrees of freedom only. Additionally, we assume the scalar profile is independent of the 3D coordinates $\rmx_{\al\gb}$. So, we set  
\begin{equation}\label{ansatz}
    W = \gO\,,\qquad C = C(\rmz, p)\,,
\end{equation}
where we introduce the variable $p := y\bar y$ and $\Omega$ is defined in \eqref{Omega:vac}. The first equation \eqref{eq:dWapp} is obviously satisfied. Substituting the ansatz \eqref{ansatz} into  \eqref{eq:dCapp}, we arrive at the following system:
\begin{subequations}
    \begin{align}
        &\p_{\rmz} C + \frac{1}{2\rmz}(p - 2\p_p - p\p^2_p)C = 0\label{eq:dCz}\\
        &C - 2\p_{p}C + \p^2_pC = 0
    \end{align}
\end{subequations}
which immediately gives rise to two branches \eqref{C:branches}
\begin{equation}\label{App:brnch}
    C(\rmz,p) =C_0^{\Delta=1}+C_0^{\Delta=2}:= \nu_1 C_1(\rmz,p) + \nu_2 C_2(\rmz,p)\,,
\end{equation}
where we denote
\begin{equation}
    C_1(\rmz,p) = \rmz e^p\,,\qquad C_2(\rmz,p) = \rmz^2(p+1)e^p\,.
\end{equation}
And now we need to check if \eqref{eq:dLapp} is satisfied. Note that $\dr_z\gO = 0$, so Eq. \eqref{eq:dLapp} becomes
\begin{equation}
    \dr_x\Lambda + \{\gO, \Lambda\}_* = 0\label{eq:dLshrt}\,.
\end{equation}
Substituting the AdS connection \eqref{Omega:vac} into \eqref{eq:dLapp} gives us
\begin{subequations}
    \begin{align}
        \p_{\rmz}\Lambda  + \frac{1}{4\rmz}[p,\Lambda]_* = 0\label{eq:dLz}\,, \\
	\dr \rmx^{\mu\nu}[y^+_{\mu}y^+_{\nu},\Lambda]_* = 0\label{eq:dLx}\,.
    \end{align}
\end{subequations}
The left-hand side of Eq. \eqref{eq:dLz} can be calculated directly. Upon integration by parts in $\tau$, the result takes the following form:
\begin{equation}
    \p_{\rmz}\Lambda + \frac{1}{4\rmz}[p,\Lambda]_* =
    \dr z^{\alpha}z_{\alpha}\int_{0}^{1}d\tau\ \tau e^{i\tau zy} \left[\p_{\rmz}C + \frac{1}{2\rmz}(\bar p C-2\p_{\bar p}C-\bar p\p^2_{\bar p}C)\right]\,,\label{eq:dLansatz}
\end{equation}
where
\begin{equation}
    \bar p:=-\tau z\bar y\quad\text{and}\quad C=C(\rmz, \bar p)\,.
\end{equation}
Notice, the integrand contains exactly Eq. \eqref{eq:dCz} with a redefined variable $\bar p$ in place of $p$. This means that both branches \eqref{App:brnch} solve \eqref{eq:dLz}. Now let us take a look at \eqref{eq:dLx}. For the  branch $\Delta=1$, $\Lambda$ acquires the following form:
\begin{equation}
    \Lambda_1 \equiv \Lambda[C_1] = \rmz\, \dr z^{\alpha}z_{\alpha} \int_{0}^{1}d\tau\ \tau e^{i\tau zy^+}\,.
\end{equation}
Note that $\Lambda_1$ depends on variables $Y$ through $y^+$ only. Now it is easy to check that \eqref{eq:dLx} is satisfied for $C_1$ corresponding to the branch $\Delta=1$. To see this, the following commutation relations appear useful:
\begin{equation}
    [y^+_{\mu}, y^+_{\nu}]_* = 0\,,\qquad [z_{\mu}, \bullet]_* = 0\,.
\end{equation}
Thus, $\Omega$ and $C_0^{\Delta=1}$ indeed solve equations \eqref{app:eqs}. One can do a similar computation to check that the $\Delta=2$ branch \textit{does not} satisfy \eqref{eq:dLx}. It is not difficult to make sure that no combination \eqref{App:brnch} other than $\nu_2=0$ satisfies \eqref{eq:dLx}. More broadly, there are no solutions of \eqref{eq:dLx} with $C=C(\rmz, p)$ except for $C\sim C_1$. Indeed, computing the commutator, Eq. \eqref{eq:dLx} boils down to 
\begin{equation}
    \dr z^{\alpha}L_{\alpha} = 0
\end{equation}
with
\begin{multline}
    \frac{1}{4}L_{\alpha} = \dr \rmx^{\mu\nu}y_{\mu}z_{\nu}z_{\alpha}\left(\int_0^1 d\tau\ \tau^2e^{i\tau zy}h - \int_0^1 d\tau\ \tau^3e^{i\tau zy}h\right) - i\dr\rmx_{\alpha}{}^{\nu}z_{\nu}\int_0^1 d\tau\ \tau^2e^{i\tau zy}h + \\  + i\dr\rmx^{\mu\nu}\bar y_{\mu}z_{\nu}z_{\alpha}\left(\int_0^1 d\tau\ \tau^2e^{i\tau zy}h - \int_0^1 d\tau\ \tau^3e^{i\tau zy}\p_p h\right)\,,\label{eq:Lcommutator}
\end{multline}
where
\begin{equation}
    h \equiv \p_p C - C\,.
\end{equation}
Each contribution from \eqref{eq:Lcommutator} has its own spinor structure, so they cannot cancel each other. Therefore we have to impose
\begin{equation}
        \int_0^1 d\tau\ \tau^2e^{i\tau zy}h = \int_0^1 d\tau\ \tau^3e^{i\tau zy}h = \int_0^1 d\tau\ \tau^3e^{i\tau zy}\p_ph = 0\,,
\end{equation}
that gives us as a consequence $h = 0$, or
\begin{equation}
    \p_p C - C = 0\,,
\end{equation}
implying the only solution of \eqref{eq:dLx} has the form
\begin{equation}
    C(\rmz, p) = f(\rmz) e^{p} \,.
\end{equation}
We see that the result is proportional to $C_1$. The function $f(\rmz)$ is determined by the vanishing of the integrand of \eqref{eq:dLansatz}.

\subsection*{Branch $\Delta = 2$}

Despite the branch $C_0^{\Delta=2}$ with $W=\Omega$ not solving \eqref{app:eqs}, it does so for the modified vacuum connection $W$. To see this, we modify the connection 1-form \eqref{Omega:vac} by adding a certain correction $\varpi$, while leaving field $C$ unchanged
\begin{equation}\label{Ansatz:w}
    W = \gO + \nu_2 \varpi(z, Y; \rmz)\,,\qquad C^{\Delta=2}_0(\rmz, p):=\nu_2C_2=\nu_2 \rmz^2(1 + p)e^p\,, 
\end{equation}
where we assume $\varpi$ to be $\nu_2$-independent. Substituting \eqref{Ansatz:w} into \eqref{app:eqs}, we have 
\begin{subequations}
    \begin{align}
        &\nu_2 (\dr_x \varpi + \{\gO,\varpi\}_*) + \nu_2^2 \varpi*\varpi = 0\,, \\
        &\nu_2(\dr_{z}\varpi + \{\gO, \Lambda_2\}_* + \dr_x\Lambda_2) + \nu_2^2\{\varpi,\Lambda_2\}_* = 0\,,\\
        &\nu_2^2(\varpi(z',y,\bar y)*_y C_2 - C_2 *_y \varpi(z',-y,\bar y))|_{z' = -y} = 0\,.
    \end{align}
\end{subequations}
As long as $\nu_2$ is an arbitrary parameter and neither of the fields $\varpi$, $\Lambda_2$, $C_2$ depend on $\nu_2$, each power of $\nu_2$ has to vanish independently. This means the following chain of equations must hold:
\begin{subequations}\label{eq:deformation}
    \begin{align}
        &\dr_x\varpi + \{\gO,\varpi\}_* = 0\,,\label{eq:dw}\\
        &\varpi*\varpi = 0\,, \label{eq:ww}\\
        &\dr_{z}\varpi + \{\gO, \Lambda_2\}_* + \dr_x\Lambda_2 = 0\,,\label{eq:dzw}\\
        &\{\varpi,\Lambda_2\}_* = 0\,, \label{eq:wL2}\\
        &(\varpi(z',y,\bar y)*_y C_2 - C_2 *_y \varpi(z',-y,\bar y))|_{z' = -y} = 0\,. \label{eq:wC2}
    \end{align}
\end{subequations}
Using the earlier results obtained in \eqref{eq:dLshrt}-\eqref{eq:dLansatz}, we note that Eq. \eqref{eq:dzw} reduces to 
\begin{equation}
    \dr_{z}\varpi = -\frac{\dr\rmx^{\al\gb}}{8\rmz}[y^+_{\al}y^+_{\gb}, \Lambda_2]_*\,.
\end{equation}
The latter equation can be solved by the standard contracting homotopy (see footnote \ref{footnote}) and we also make use of  \eqref{eq:Lcommutator} for the commutator on the r.h.s. Specifically, we find
\begin{equation}\label{w:corr}
    \varpi = \frac{z^{\al}}{8\rmz}\int_0^1 d\tau L_{\al}(\tau z, y) = \frac{i\rmz}{2}\dr\rmx^{\al\gb}z_{\al}z_{\gb}\int_0^1d\tau\  \tau(1-\tau)e^{i\tau zy^+}\,.
\end{equation}
Now we need to check if the obtained deformation $\varpi$ satisfies the whole chain of equations \eqref{eq:deformation}. The following calculation allows us to see that this is indeed the case. First off
\begin{equation}
    z_{\al}z_{\gb} e^{it zy^+} * z_{\mu}z_{\nu} e^{i\tau zy^+} = \frac{(1 - t)^2(1-\tau)^2}{(1 - \tau t)^6}\ z_{\mu}z_{\nu}z_{\alpha} z_{\beta}\exp\left[izy^+\frac{t+\tau-2\tau t}{1 - \tau t}\right] \label{eq:ExOfStarProduct}\,,
\end{equation}
which immediately proves \eqref{eq:ww} because
\begin{equation}
    \varpi * \varpi \sim \dr\rmx^{\al\gb} \wedge \dr\rmx^{\mu\nu} z_{\alpha}z_{\beta}z_{\mu}z_{\nu} = 0\,.
\end{equation}
The analysis of \eqref{eq:wL2}, \eqref{eq:wC2} reveals similar to \eqref{eq:ExOfStarProduct} star products, as we find that both equations are indeed satisfied. That the first equation \eqref{eq:dw} is satisfied can be observed from comparing $\varpi$ and $\Lambda_0=\nu_1\Lambda_1$ from \eqref{Lambda0},
\begin{equation}
    \varpi = \frac{i\rmz}{2}\dr\rmx^{\al\gb}\int_0^1d\tau(1-\tau)\tau\,z_{\al}z_{\gb}e^{i\tau zy^+}\quad \text{vs}\quad \Lambda_1 = \rmz\, \dr z^{\alpha} \int_{0}^{1}d\tau\tau\, z_{\alpha}e^{i\tau zy^+}\,.
\end{equation}
It is important that both $\varpi$ and $\Lambda_1$ have the same dependence on $\rmz$. It can be seen that equations \eqref{eq:dw} and \eqref{eq:dLshrt} are essentially the same equations; see also the results of the previous subsection. Namely, it is easy to see that both Eq. \eqref{eq:dLz} and Eq. \eqref{eq:dLx} are satisfied for $\varpi$ for the same reasons as for $\Lambda_1$. This means that the whole chain of equations \eqref{eq:deformation} is satisfied. Eventually, we see that the fields $W=\gO + \nu_2 \varpi$ and $C=\nu_2 C_2$ form a solution. Now that it is shown that the two branches satisfy equations of motion separately, a question arises whether an arbitrary mixture of these with coefficients $\nu_1$ and $\nu_2$ is still a vacuum? This would be the case provided 
\begin{equation}
    [\varpi, \Lambda_1]_*=0\,,
\end{equation}
which is straightforward to verify. Therefore, \eqref{W:vac} and \eqref{C:vac} solve \eqref{Eqs}.

\renewcommand{\theequation}{\Alph{appendix}.\arabic{equation}}
\addtocounter{appendix}{1} \setcounter{equation}{0}
\addtocounter{section}{1}
\addcontentsline{toc}{section}{\,\,\,\,\, B. Vertex integration domains}

\section*{B. Vertex integration domains}\label{app:B}
Here we provide a proof of the important property of the integration domains
\begin{equation}
    \bigsqcup_{k-\textnormal{odd}} \mathcal{D}^{[k]}_n = \bigsqcup_{k-\textnormal{even}} \mathcal{D}^{[k]}_n\,,\label{eq:DomainProperty}
\end{equation}
where the domains $\mathcal{D}_{n}^{[k]}$ appear in the construction of the higher-spin interaction vertices \eqref{verC}. These are defined by a set of inequalities:
\begin{equation}
    \label{eq:VertexDomainApp}
    \mathcal{D}^{[k]}_n =
    \begin{cases}
        \eta_1 + ... + \eta_{n} = 1\,,\quad\eta_i\geq 0\,, \\
        \xi_1 + ... + \xi_{n} = 1\,,\quad\xi_i\geq 0\,, \\
        C_i \leqslant 0 \,, \quad i \in [1 \,,\, k-1]\,, \\
        C_i \geqslant 0\,, \quad i \in [k+1\,,\,n-1]\,.
    \end{cases},\ C_i \equiv \eta_i \xi_{i+1} - \eta_{i+1} \xi_{i}
\end{equation}
or in a short form as follows
\begin{equation}
    \mathcal{D}^{[k]}_n = \{C_1 \leqslant 0, ..., C_{k-1}\leqslant 0|C_{k+1} \geqslant 0, ..., C_{n}\geqslant 0 \}\,.
\end{equation}
Let us introduce a set of auxiliary domains
\begin{equation}
    \Delta_n^{[k]} \equiv \{C_1 \leqslant 0,\dots, C_{k-1}\leqslant 0|C_{k} \geqslant 0,\dots, C_{n}\geqslant 0 \}\,,\quad \Delta_n^{[1]} \equiv \mathcal{D}^{[0]}_n\,,\quad \Delta_n^{[n]} \equiv \mathcal{D}^{[n]}_n\,.
\end{equation}
Here all the assumptions for $\xi, \eta$ are the same as for $\mathcal{D}_n^{[k]}$. Using these domains, we can rewrite $\mathcal{D}$ as
\begin{equation}
    \mathcal{D}_n^{[k]} = \Delta_n^{[k]} \cup \Delta_n^{[k+1]}\,.
\end{equation}
Intersection $\Delta_n^{[k]}\cap\Delta_n^{[k+1]}$ is not empty but a set of zero measure:
\begin{equation}
    \Delta_n^{[k]}\cap\Delta_n^{[k+1]} = \{C_1 \leqslant 0,\dots, C_{k-1}\leqslant 0|C_{k} = 0 | C_{k+1} \geqslant 0,\dots, C_{n}\geqslant 0 \}
\end{equation}
Recall that we are dealing with integration domains, so this intersection can be freely omitted. To emphasize this we use $\sqcup$ as union up to a set of measure zero. Now one easily obtains the property \eqref{eq:DomainProperty}
\begin{equation}
    \bigsqcup_{k-\textnormal{odd}} \mathcal{D}^{[k]}_n = \Delta_n^{[1]} \sqcup \Delta_n^{[2]} \sqcup ... \sqcup \Delta_n^{[n-1]} \sqcup \Delta_n^{[n]} = \bigsqcup_{k-\textnormal{even}} \mathcal{D}^{[k]}_n\,.
\end{equation}

\renewcommand{\theequation}{\Alph{appendix}.\arabic{equation}}
\addtocounter{appendix}{1} \setcounter{equation}{0}
\addtocounter{section}{1}
\addcontentsline{toc}{section}{\,\,\,\,\, C. {Vertices $\Upsilon(\w(y^+), C_0^n)$}}

\section*{C. Vertices $\Upsilon(\w(y^+), C_0^n)$}

Let us show that the vertices $\Upsilon(\w(y^+), C_0^n)$, while emerging in  the linearization around \eqref{vac}, appear to be zero for any $n$. To this end, we rewrite vertex \eqref{verC} in the differential (Taylor) form using the notation from \cite{Didenko:2018fgx}, \cite{Didenko:2024zpd} as 
\begin{equation}\label{eq:verCapp}
    \Upsilon(\omega, C^n) =
    \sum_{k=0}^n \Phi^k_n(y|t;p)\left(\prod_{i = 1}^{k} \star C_0(y_i, \bar y)\right) \star \omega(y_t,\bar y) \star \left(\prod_{i = k + 1}^{n} \star C_0(y_i, \bar y)\right) \Big|_{y_i = y_t = 0}\,,
\end{equation}    
where 
\begin{equation}
    \Phi^k_n(y|t;p) = (-)^{k+1}(ty)^{n-1}\smashoperator{\int_{\mathcal{D}_{n}^{[k]}}}e^{-iyP_n(\xi)-itP_n(\eta)+ityS^k_n}
\end{equation}
contains the following differential operators
\begin{equation}
         t_{\mu} = -i \frac{\p}{\p y_t^{\mu}}\,,\qquad p_{i\mu} = -i \frac{\p}{\p y_i^{\mu}}
\end{equation}
with 
\begin{equation}
    S^{[k]}_n = -\sum_{s=1}^{k} \xi_s + \sum_{s=k+1}^{n} \xi_s + \sum_{i<j}^{n}(\xi_i \eta_j - \xi_j \eta_i)\,,
\end{equation}
and 
\begin{equation}
     P_n(a) = \sum_{i = 0}^n p_i a_i\,.
\end{equation}
We need now to calculate $\star$ products of $C_0$. Additionally, we set $\nu=1$ for convenience, since its particular value is not important for the following analysis. For the product of two $C_0$'s one easily obtains
\begin{equation}
    C_0(y_1, \bar y) \star C_0(y_2, \bar y) = \rmz^2 e^{y_1 \bar y} \star e^{y_2 \bar y} = \rmz^2 e^{(y_1 + y_2) \bar y + iy_1y_2}\,.
\end{equation}
Therefore,
\begin{equation}
    \prod_{i = 1}^{k} \star C_0(y_i, \bar y) = \rmz^k \exp\left[\sum_{s=1}^k y_s \bar y + i\sum_{1\leq i < j\leq k} y_i y_j \right]\,.
\end{equation}
Similarly,
\begin{multline}
    \left(\prod_{i = 1}^{k} \star C_0(y_i, \bar y)\right) \star \omega(y',\bar y) \star \left(\prod_{i = k + 1}^{n} \star C_0(y_i, \bar y)\right) = \\
    = \rmz^n \exp\left[\sum_{s=1}^n y_s \bar y + i\sum_{1\leq i < j\leq n} y_i y_j \right] \omega\left(y',\bar y - i \sum_{s=1}^k y_s + i\sum_{s=k+1}^n y_s \right)\,.
\end{multline}
Substituting this into \eqref{eq:verCapp} and using that 
\begin{equation}
\sum_{i=1}^n \xi_i = \sum_{i=1}^n \eta_i = 1   
\end{equation}
 gives us
\begin{multline}
    \Upsilon(\omega, C_0^n) = \sum_{k=0}^n (-)^{k+1}(ty)^{n-1}\smashoperator{\int_{\mathcal{D}_{n}^{[k]}}} \exp\left[(y + t)\bar y + i(ty)\left(1 - 2\sum_{1}^k \xi_s \right)\right] \times \\
    \times \omega\left[y_t,\bar y + iy\left(1 - 2\sum_1^k \xi_s\right) + it\left(1 - 2\sum_1^k\eta_s \right) \right]\Big|_{y_t = 0}\,.
\end{multline}
Now we substitute $\omega(y, \bar y) = \w(y + i\bar y)$ to arrive at
\begin{equation}
    \Upsilon(\w, C_0^n)
    = e^{y\bar y} \sum_{k=0}^n (-)^{k+1} (ty)^{n-1} \smashoperator{\int_{\mathcal{D}_{n}^{[k]}}} \w\left[y_t- t\left(1 - 2\sum_1^k\eta_s \right) \right]\Big|_{y_t = 0}\,.
\end{equation}
As a final step, in order to act with operator $t$ on $y_t$ from the left, we use the decomposition 
\begin{equation}
    \w(y' - A t) = \int d^2u d^2v\ e^{i u (v + At)} \w(y' + v)\,,
\end{equation}
which upon action of the translation operator generated by the exponential of $t$ takes us to the following result:
\begin{equation}
    \Upsilon(\w(y^+), C_0^n) = e^{y\bar y} (ty)^{n-1} \sum_{k=0}^n (-)^{k+1} \smashoperator{\int_{\mathcal{D}_{n}^{[k]}}} \w(y_t)|_{y_t = 0}\,. 
\end{equation}
The latter expression is just zero as a consequence of \eqref{domain}, thus,
\begin{equation}
    \Upsilon(\w(y^+), C_0^n) =0\,.
\end{equation}

\renewcommand{\theequation}{\Alph{appendix}.\arabic{equation}}
\addtocounter{appendix}{1} \setcounter{equation}{0}
\addtocounter{section}{1}
\addcontentsline{toc}{section}{\,\,\,\,\, D. Consistency check}

\section*{D. Consistency}
Here we check consistency of the linearized equations \eqref{Tx:nonconsv}, \eqref{Tz:nonconsv}, which we can write down as %{\color{red} check $\nu$ factor} 
\begin{subequations}
    \begin{align}
        &\p_{\rmz} T = \frac{1}{2}(\p\bar \p T
        + \nu w\bar \p J[T])\label{eq:linTappa}\,,\\
        &\dr_{\bold{x}}T = \frac{i}{2} \dr x^{\alpha\beta} (\p_{\alpha}\bar\p_{\beta}T + \nu w_{\alpha}\bar\p_{\beta} J[T])\,,\label{eq:linTappb}
    \end{align}\label{eq:linTapp}
\end{subequations}
Specifically, the following two integrability conditions should be satisfied: 
\begin{equation}
    \dr_{\rmx}^2 T = 0\,,\qquad [\dr_{\rmx}, \p_{\rmz}]T = 0\,. 
\end{equation}
Let us introduce also the following shorthand notation:
\begin{align}
    &T^- = T(\tau w, \bar{w} - i[1-\tau] w)\,,\quad T^+ = T(\tau w, \bar{w} + i[1-\tau] w)\,,\\
    &J[T] = \int_0^1 d\tau (1 - \tau) (T^- - T^+)\,,\\
    &H_{\mu\nu}[T] = \frac{1}{4}\left(\p_{\mu}\bar\p_{\nu}T + \p_{\nu}\bar\p_{\mu}T + \nu w_{\mu}\bar\p_{\nu}J[T] + \nu w_{\nu}\bar\p_{\mu}J[T]\right)\,,
\end{align}
where we recall that spinorial $\partial$ and $\bar\partial$ denote differentiation with respect to the full arguments of $T(s, \bar s)$. When necessary, we distinguish differentiation with respect to a particular variable by marking  
\begin{equation}\label{primedvariables}
    w' = w'' = \tau w\,,\quad \bar w' = \bar w - i(1 - \tau)w\,,\quad \bar w'' = \bar w + i(1 - \tau)w\,,
\end{equation}
which brings in our calculations differentiations of the form 
\begin{equation}\label{primeddiffs}
    \p' = \frac{\p}{\tau} + i\frac{1 - \tau}{\tau}\bar\p,\ \p'' = \frac{\p}{\tau} - i\frac{1 - \tau}{\tau}\bar\p\,.
\end{equation}
Let us start with checking the consistency of \eqref{eq:linTappb}.
\begin{itemize}
\item Equation \eqref{eq:linTappb} is rewritten as
\begin{equation}
    \dr_{\rmx}T = i \dr \rmx^{\mu\nu}H_{\mu\nu}[T]\,,
\end{equation}
while its consistency condition is
\begin{equation}
    0 = i \dr \rmx^{\alpha\beta}\wedge \dr\rmx^{\mu\nu}\p_{\alpha\beta}H_{\mu\nu}[T]\,.
\end{equation}
The $\nu$-independent contribution cancels out due to consistency of the $\nu = 0$ case. We are left with the following:
\begin{equation}
    \begin{split}
        &4 \dr \rmx^{\alpha\beta}\wedge \dr \rmx^{\mu\nu} \p_{\alpha\beta}H_{\mu\nu}[T] = i \nu\dr \rmx^{\alpha\beta}\wedge \dr \rmx^{\mu\nu}\bar\p_{\nu}\bar\p_{\beta} (w_{\alpha}\p_{\mu} + \epsilon_{\mu\alpha})J[T] + \\
        &+ 2\nu \dr \rmx^{\alpha\beta}\wedge \dr \rmx^{\mu\nu} w_{\mu}\bar\p_{\nu}\int_0^1 d\tau (1 - \tau) \p_{\alpha\beta}(T^- - T^+)\,.
    \end{split}
\end{equation}
The integrand is further transformed using \eqref{primedvariables}, \eqref{primeddiffs}
\begin{equation}
    \begin{split}
        &\p_{\alpha\beta} T^- = \frac{i}{4}(\p'_{\alpha}\bar\p'_{\beta} + \p'_{\beta}\bar\p'_{\alpha})T(w', \bar w') + \frac{i\nu}{4} w'_{\alpha}\bar\p'_{\beta}J[T(w', \bar w')] = \\ 
        &= \frac{i}{4\tau}(\p_{\alpha} \bar\p_{\beta} + \p_{\beta} \bar\p_{\alpha}) T^- + i \frac{1 - \tau}{2\tau}\bar\p_{\alpha} \bar\p_{\beta}T^- + \frac{i\nu\tau}{4} (w_{\alpha}\bar\p_{\beta} + w_{\beta}\bar\p_{\alpha}) J[T^-]\,.
    \end{split}
\end{equation}
The last term above brings the $\nu^2$ terms, which cancel independently
\begin{equation}
    \nu^2 \cdot \dr \rmx^{\alpha\beta}\wedge \dr \rmx^{\mu\nu} w_{\mu}\bar\p_{\nu}w_{\alpha}\bar\p_{\beta}\int_0^1d\tau(1 - \tau)\tau J[T^-] = 0
\end{equation}
and analogously with $T^+$. So, one is only left to check the terms proportional to $\nu$
\begin{equation}
    \begin{split}
        &-\frac{4i}{\nu} \dr \rmx^{\alpha\beta}\wedge \dr \rmx^{\mu\nu}\p_{\alpha\beta}H_{\mu\nu} = \dr \rmx^{\alpha\beta}\wedge \dr \rmx^{\mu\nu} \bar\p_{\nu}\bar\p_{\beta}(w_{\alpha}\p_{\mu} + \epsilon_{\mu\alpha})J[T] + \\ 
        &+ \dr \rmx^{\alpha\beta}\wedge \dr \rmx^{\mu\nu} \bar\p_{\nu}\bar\p_{\beta}\int_0^1d\tau (1 - \tau) \left[ \frac{1}{\tau} w_{\mu}\p_{\alpha} T^- + i \frac{1 - \tau}{\tau} w_{\mu}\bar\p_{\alpha} T^- \right] - \\ 
        &- \dr \rmx^{\alpha\beta}\wedge \dr \rmx^{\mu\nu} \bar\p_{\nu}\bar\p_{\beta}\int_0^1d\tau (1 - \tau) \left[ \frac{1}{\tau} w_{\mu}\p_{\alpha} T^+ - i \frac{1 - \tau}{\tau} w_{\mu}\bar\p_{\alpha} T^+ \right]\,.
    \end{split}
\end{equation}
Now we use the antisymmetrization rule for two-component spinors
\begin{equation}
    2\dr \rmx^{\alpha\beta} \wedge \dr \rmx^{\mu\nu}\bar\p_{\nu}\bar\p_{\beta} = \epsilon^{\alpha\mu} \cdot \dr \rmx_{\rho}^{\hspace{5pt}\beta} \wedge \dr \rmx^{\rho\nu}\bar\p_{\nu}\bar\p_{\beta}
\end{equation}
and the integration by parts using
\begin{equation}
    \begin{split}
        &\frac{\p}{\p\tau}T^- = w^{\mu}(\p'_{\mu} + i \bar\p'_{\mu})T^- = \frac{w^{\mu}(\p + i \bar\p)_{\mu}}{\tau}T^-\,, \\
        &\frac{\p}{\p\tau}T^+ = w^{\mu}(\p'_{\mu} - i \bar\p'_{\mu})T^+ = \frac{w^{\mu}(\p - i \bar\p)_{\mu}}{\tau}T^+\,.
    \end{split}
\end{equation}
The result is
\begin{equation}
    \epsilon_{\alpha\mu}\dr\rmx^{\alpha\beta} \wedge \dr\rmx^{\mu\nu}\bar\p_{\nu}\bar\p_{\beta} [T(0; \bar w + iw) - T(0; \bar w - iw)] = 0\,.
\end{equation}
Which is satisfied by the following general $T$-module
\begin{equation}\label{app:Tcons}
    T(0, \bar w) = T_0 + T_{\alpha}\bar w^{\alpha} + \frac{1}{2}T_{\alpha\beta}\bar w^{\alpha}\bar w^{\beta}\,.
\end{equation}

\item Now we need to check the condition $[\dr_{\rmx}, \p_{\rmz}]T = 0$. The corresponding equation splits into three with respect to power of $\nu$. Again, the $\nu$-independent part vanishes automatically and we are left with the following:
\begin{equation}
    \begin{split}
        &\nu:\quad \dr \rmx^{\alpha\beta}(\p_{\alpha}\bar\p_{\beta} w\bar\p J[T] - \p\bar\p w_{\alpha}\bar\p_{\beta} J[T]) + \\
        &+ \dr \rmx^{\alpha\beta} w_{\alpha}\bar\p_{\beta} \int_0^1 d\tau(1 - \tau)[\p'\bar\p T^- - \p''\bar\p T^+] - \\
        &- \dr \rmx^{\alpha\beta}w\bar\p \int_0^1 d\tau(1 - \tau)[\p'_{\alpha}\bar\p_{\beta} T^- - \p''_{\alpha}\bar\p_{\beta} T^+] = 0\,,
    \end{split}\label{eq:dxdznu}
\end{equation}
\begin{equation}
    \begin{split}
        &\nu^2:\quad \dr \rmx^{\alpha\beta}w_{\alpha}\bar\p_{\beta} \int_0^1 d\tau(1 - \tau)(w'\bar\p J[T^-] - w''\bar\p J[T^+]) - \\
        &- \dr\rmx^{\alpha\beta} w\bar\p \int_0^1 d\tau(1 - \tau) (w'_{\alpha}\bar\p_{\beta} J[T^-] - w''_{\alpha}\bar\p_{\beta} J[T^+]) = 0\,.
    \end{split} \label{eq:dxdznu2}
\end{equation}
Recalling that $w' = w'' = \tau w$, one can see that equation \eqref{eq:dxdznu2} is satisfied. To check the first equation \eqref{eq:dxdznu} one need to use the following useful feature of two-dimensional spinors
\begin{equation}
    w\bar\p\p_{\alpha}\bar\p_{\beta} - w_{\alpha}\bar\p_{\beta}\p\bar\p = \bar\p_{\beta}\bar\p^{\gamma}(w_{\gamma}\p_{\alpha} - w_{\alpha}\p_{\gamma}) = \epsilon_{\gamma\alpha} \cdot \bar\p_{\beta}\bar\p^{\gamma} w\p\,.
\end{equation}
Using this expression and integrating by parts \eqref{eq:dxdznu} one arrives at the same result
\begin{equation}
    \bar\p_{\alpha}\bar\p_{\beta} [T(0; \bar w + iw) - T(0; \bar w - iw)] = 0\,.
\end{equation} 

%{\color{red} don't forget to adjust the text below}
Constraint \eqref{app:Tcons} should commute with $\dr_\rmx$ and $\dr_\rmz$, thus leading to a set of additional constraints. These turn out to be the following:   
\begin{equation}
    T_{\alpha(m), \beta(n)} = 0\quad\text{for }\quad n - m > 2\,.
\end{equation}
To prove this, consider our equations \eqref{Tx:nonconsv}, \eqref{Tz:nonconsv}, which we combine in a single one  (recall $x=(\rmx, \rmz)$)
\begin{equation}\label{app:dxT}
    \dr_{x}T_{\al(m), \gb(n)} = -i\rmz \e^{\gamma\rho}\left(T_{\gamma\alpha(m),\rho\beta(n)} + 4i\nu \epsilon_{\alpha\gamma} \sum_{k = 1}^{\lfloor \frac{m}{2} \rfloor} \frac{k(-)^{k}}{m + 1}T_{\alpha(m - 2k),\alpha(2k - 1)\beta(n)\rho}  \right)
\end{equation}
using the vierbein \eqref{vierbein}. The consistency condition \eqref{app:Tcons} reads
\begin{equation}
    \bar j_{\beta(n)} : = T_{\alpha(0), \beta(n)} = 0\,,\qquad n > 2\label{consistencyApp}
\end{equation}
For the cases where  $m = 0, 1$, the sum on the right of \eqref{app:dxT} does not contribute. We have 
\begin{subequations}
    \begin{align}
        &\dr_{x}\bar j_{\gb(n)} = -i\rmz \e^{\gamma\rho}T_{\gamma,\rho\beta(n)}\,,\label{zeromComp}\\
        &\dr_{x} T_{\alpha,\gb(n)} = -i\rmz \e^{\gamma\rho}T_{\gamma\alpha,\rho\beta(n)}\,.\label{onemComp}
    \end{align}
\end{subequations}
Combining together \eqref{consistencyApp}, \eqref{zeromComp} and \eqref{onemComp}, we conclude that
\begin{equation}
    T_{\alpha, \beta(n + 1)} = 0,\ T_{\alpha(2), \beta(n + 2)} = 0\quad \text{for all}\quad n > 2\,.
\end{equation}
Notice that the above equations hold true because the vierbein contains both symmetric and antisymmetric parts of spinorial indices. Now we return to the general equation \eqref{app:dxT} and rewrite it as follows:
\begin{equation}
    i\rmz \e^{\gamma\rho}T_{\gamma\alpha(m),\rho\beta(n)} = -\dr_{x}T_{\al(m), \gb(n)} + 4\rmz \e^{\gamma\rho} \nu \epsilon_{\alpha\gamma} \sum_{k = 1}^{\lfloor \frac{m}{2} \rfloor} \frac{k(-)^{k}}{m + 1}T_{\alpha(m - 2k),\alpha(2k - 1)\beta(n)\rho}\,.
\end{equation}
We now look at the $n - m > 2$ case. Suppose we have already proven that 
\begin{equation}
    T_{\alpha(m), \beta(n > m + 2)} = 0\quad \text{for all}\quad m, m-1, m-2,\dots, 0\label{AssumptionApp}
\end{equation}
For any $m$ and $n$ the sum consists of components $T_{\alpha(m - 2k), \alpha(2k - 1)\beta(n)\rho}$ whose difference of number of indices is $n - m + 4k > n - m > 2$. This means that for any $k$ this component is covered by the assumption \eqref{AssumptionApp}. We conclude then
\begin{equation}
    \e^{\gamma\rho}T_{\gamma\alpha(m),\rho\beta(n)} = 0\quad \Rightarrow\quad T_{\alpha(m + 1),\beta(n + 1)} = 0\quad \text{for all}\quad n - m > 2\,,
\end{equation}
thus completing the proof.

\end{itemize}

\renewcommand{\theequation}{\Alph{appendix}.\arabic{equation}}
\addtocounter{appendix}{1} \setcounter{equation}{0}
\addtocounter{section}{1}
\addcontentsline{toc}{section}{\,\,\,\,\, E. A remark on star-periodic functions}

\section*{E. A remark on star-periodic functions}
This Appendix is aimed at demonstrating that functions of the form 
\begin{equation}\label{exp:D}
    \exp_*{\tau D}:=\sum_{n=0}^{\infty}\frac{\tau^n}{n!}\underbrace{D*\dots *D}_{n}\,,
\end{equation}
where the dilatation operator $D$ is defined in \eqref{D}, may not have well-defined symbols for certain values of $\tau$. In particular, this is the case for periodic functions with $\tau=2\pi in$, where $n$ is an integer. 

Suppose $\tau$ is an arbitrary number and let us look for a symbol of the exponential \eqref{exp:D} in the following form: 
\begin{equation}
    \exp_*{\tau D}=h(\tau)\exp{(q(\tau)D)}\,,
\end{equation}
where $h(\tau)$ and $q(\tau)$ are yet to be determined. Star multiplying \eqref{exp:D} with $D$ we have, on the one hand, 
\begin{equation}\label{eq:der}
    D*\exp_*{\tau D}=\frac{\partial}{\partial\tau}\exp_*{\tau D}=(h'(\tau)+h(\tau)q'(\tau)D)\exp{(q(\tau) D)}\,,
\end{equation}
where $f'(\tau)$ denotes derivative with respect to the variable $\tau$. On the other hand, using \eqref{star:gen} one finds that 
\begin{equation}
    D*=-\frac{1}{4}(y+i\partial)_{\al}(\bar y+i\bar\partial)^{\alpha}\,,
\end{equation}
from which it is straightforward to obtain
\begin{equation}\label{eq:str}
   D*\exp_*{\tau D}:=D*\left(h \exp{q D}\right)=\left(D-\frac{1}{8}q-\frac{1}{16}q^2 D\right)h\exp{q D}\,. 
\end{equation}
Equating the right-hand sides of \eqref{eq:der} and \eqref{eq:str} to each other gives us the following differential equations on functions $h$ and $q$:
\begin{align}\label{eq:hq}
    &h'=-\ff18 h\cdot q\,,\\
    &q'=1-\left(\ff{q}{4}\right)^2\,.
\end{align}
For $\tau=0$, in addition, we have $\exp_*(0)=1$. This brings us to the solution of the form
\begin{align}
    q(\tau) = 4\tanh{\frac{\tau}{4}}\,,\quad  h(\tau) = \frac{1}{\cosh^2{\frac{\tau}{4}}}
\end{align}
Therefore, for the symbol of the star-exponential, we have
\begin{equation}\label{eq: expD}
    \exp_*(\tau D)=\frac{1}{\cosh^2{\frac{\tau}{4}}}\exp{ \left(4D\tanh{\frac{\tau}{4}}\right)}\,,
\end{equation}
or changing parameterization by introducing $s:=e^{\tau/2}$,
\begin{equation}
    \exp_*(D\log s^2)=\frac{4 s}{(s+1)^2}\exp{4\left(\frac{s-1}{s+1}\right)D}\,.
\end{equation}
Notably, for $\tau=2\pi i n$ from \eqref{eq: expD} it follows that
\begin{equation}\label{eq:weird}
    \exp_*{(2\pi i n D) }=
    \begin{cases}
        1\,,\quad n -\text{even}\\
        \infty\,, \quad n-\text{odd}
    \end{cases}
\end{equation}
$n\in\mathrm{Z}$ is exactly the case that has emerged in the process of analyzing periodic degrees of freedom in Section \ref{sec:lin}. This result indicates that for special values of $n$, such as in \eqref{eq:weird}, the star exponential \eqref{exp:D} is not analytic in $D$. Perhaps, it may make sense as a distribution.

\end{document}